\documentclass[12pt]{nature}
\usepackage{amsmath,amsfonts,amssymb,amsthm,array}
\usepackage{dsfont,color}
\usepackage{microtype} %%% To avoid \hbox overfull
\usepackage[final]{pdfpages}
\usepackage{braket}
\usepackage{float} % for [H] option
\usepackage[colorlinks,citecolor=blue,linkcolor=blue,urlcolor=blue,hypertexnames=false]{hyperref}
\newcommand{\figref}[2]{\hyperref[#1]{\ref*{#1}#2}}

\usepackage{hyperref}
\usepackage{doi}

\usepackage[titletoc,title]{appendix}
\makeatletter 
\let\saved@includegraphics\includegraphics
\AtBeginDocument{\let\includegraphics\saved@includegraphics}
\renewenvironment{figure}{\@float{figure}}{\end@float}
\makeatother

\newcommand{\etal}{\textit{et al}. }
\newcommand{\ie}{\textit{i}.\textit{e}., }

\title{
\textcolor{black}{
Magnetochiral Charge Pumping due to Charge Trapping and Skin Effect in Chirality-Induced Spin Selectivity}
}

\author{
\center Yufei Zhao$^{1\ast}$, Kai Zhang$^{2}\footnote{These authors contribute equally.}$, Jiewen Xiao$^1$, Kai Sun$^2$, and  Binghai Yan $^{1,3}$ \footnote{binghai.yan@weizmann.ac.il} \\
\normalsize{
$^1$ Department of Condensed Matter Physics, Weizmann Institute of Science, Rehovot 76100, Israel \\
$^2$ Department of Physics, University of Michigan, Ann Arbor, MI 48109, USA\\
$^3$ Department of Physics, the Pennsylvania State University, University Park, PA, 16802, USA
}}

\begin{document} 

\maketitle 

\begin{abstract}
\textcolor{black}{
Chirality-induced spin selectivity (CISS) generates giant spin polarization in transport through chiral molecules, paving the way for novel spintronic devices and enantiomer separation. Unlike conventional transport, CISS magnetoresistance (MR) violates Onsager’s reciprocal relation, exhibiting significant resistance changes when reversing electrode magnetization at zero bias. However, its underlying mechanism remains unresolved.
In this work, we propose that CISS MR originates from charge trapping that modifies the electron tunneling barrier and circumvents Onsager's relation, distinct from previous spin polarization-based models. Charge trapping is governed by the non-Hermitian skin effect, where dissipation leads to exponential wavefunction localization at the ferromagnet-chiral molecule interface. Reversing magnetization or chirality alters the localization direction, changing the occupation of impurity/defect states in the molecule (\textit{i.e.}, charge trapping) -- a phenomenon we term magnetochiral charge pumping.
Our theory explains why CISS MR can far exceed the ferromagnet spin polarization and why chiral molecules violate the reciprocal relation but chiral metals do not.
Furthermore, it predicts exotic phenomena beyond the conventional CISS framework, including asymmetric MR induced by magnetic fields alone (without ferromagnetic electrodes), as confirmed by recent experiments. This work offers a deeper understanding of CISS and opens avenues for controlling electrostatic interactions in chemical and biological systems through the magnetochiral charge pumping.
}

% The interplay between chirality, spin, and electron transport reveals unexpected quantum mechanical phenomena that challenge traditional transport theories. For example, chiral molecules like DNA generate giant spin polarization in nanodevices characterized by large magnetoresistance (MR). This phenomenon, called chirality-induced spin selectivity (CISS), paves a pathway for unconventional spintronic devices and enantiomer separation. Distinct from ordinary transport, CISS MR violates Onsager's reciprocal relation, and its physical mechanism remains elusive. In this work, we propose that the CISS MR originates from the charge trapping that modifies the electron tunneling barrier and circumvents Onsager's reciprocity, which is indirectly related to spin polarization. Our theory reveals the intimate connection between CISS MR and non-Hermitian skin effect in which the dissipation induces exponential localization of wave functions at system boundaries. 
% The localization direction influences the impurity state occupation near the ferromagnet-molecule interface and thus modulates the charge trapping. We provide an accessible tunneling barrier model to extract the barrier modulation from available experimental results. Our work provides a deep understanding of the CISS transport and insights to explore the magnetochiral interaction regarding spin, charge and chirality in chemical and biological interactions. 
\end{abstract}

\section{Introduction}

Chirality is a fundamental concept in chemistry, physics, and biology~\cite{kelvin1894molecular}. Recently, chirality was reported to generate spin polarization that is far larger than an ordinary ferromagnet (FM), called chiral induced spin selectivity (CISS)~\cite{Gohler2011}. CISS opens exciting pathways to design unconventional spintronic devices using chiral molecules, perform chiral separation via magnetism, and explore the spin-selective biological process~\cite{Naaman2012,Naaman2019}. 
However, the physical mechanism underlying CISS is elusive and debated~\cite{evers2021theory,yan2024structural}, although some spin polarization-based models~\cite{Gersten2013,liu2021chirality,das2022temperature,hedegaard2023spin}
%, for example, electron-phonon coupling~\cite{das2022temperature} or metal-molecule interface effects~\cite{liu2021chirality,Dubi2021,hedegaard2023spin}, 
reflect some consistent features with experiments.

\begin{figure}[p]
    \centering
    \includegraphics[width=1\linewidth]{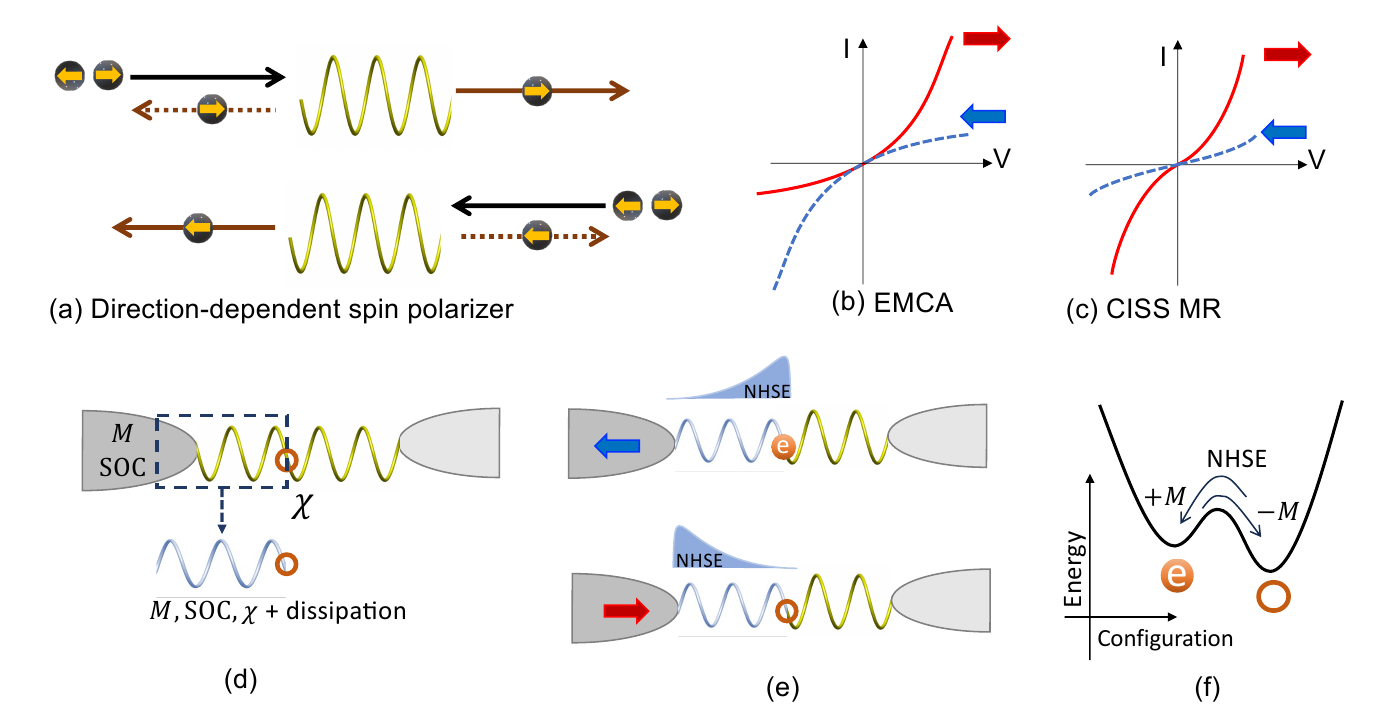}
    \caption{
    \textbf{Chirality-driven magnetoresistance (MR) including the electrical magnetochiral anisotropy (EMCA) and chirality-induced spin selectivity (CISS) .} (a) Transmitted (solid line) and reflected (dashed line) electrons through a chiral molecule get spin-polarized (indicated by small yellow arrows) for unpolarized incident electrons. In a perturbative picture, because the spin polarization relies on the incident direction, the resistance is direction-dependent when the spin polarizer is connected to a ferromagnetic electrode (magnetization indicated by the red/blue arrows), leading to EMCA rather than CISS MR. (b) and (c) show the typical I-V curves for EMCA and CISS MR, respectively. The violation of Onsager's relation is characterized by the change of zero-bias conductance upon switching the electrode magnetization. (d) In a CISS device, the ferromagnet-molecule interface exhibits magnetization($M$), spin-orbit coupling (SOC), chirality ($\chi$), and dissipation. A chiral chain model (light blue) is adopted to represent the interface with coexisting $M$, SOC, $\chi$ and and dissipation. (e) At the interface, the wave function is exponentially localized to one side due to the non-Hermitian skin effect (NHSE) when current flows. Merely NHSE leads to EMCA -- another interpretation of EMCA besides the spin polarization. If an impurity state(circle) exists on the molecule side, the asymmetric wave function due to NHSE can control this state occupied (electron-trapping) or empty. (f) Schematics of energy profile for the electron-trapping state and no trap state as two metastable phases. NHSE drives the switch between two phases by reserving $M$ (or $\chi$), \ie the magnetochiral charge pumping.
    }
    \label{fig:device}
\end{figure}

In experiments, CISS is commonly probed by a two-terminal  magnetoresistance (MR) ~\cite{Xie2011,Kiran2016,Varade2018,Liu2020,Kim2021,Qian2022,Liu2022experiment,Al-Bustami2022} where the chiral molecule is sandwiched between an FM electrode and a nonmagnetic lead (see Fig.~\figref{fig:device}{d} for illustration). The resistance changes as flipping the electrode magnetization ($M$). Because MR exhibits a feature similar to a magnetic tunneling junction, chiral molecules were usually presumed to induce spin polarization. The CISS MR increases with bias and molecule length and can be even larger than 60\% in reports~\cite{Kulkarni2020,Mishra2020,Al-Bustami2022} although the FM electrode such as Ni exhibits only 20\% spin polarization. 
Despite that tremendous theoretical efforts~\cite{evers2021theory} were made to interpret the induced spin polarization by chiral molecules, it is still impossible to rationalize the giant CISS MR ratio by assuming even 100\% spin polarization in the molecule and matching it to the FM contact (20\%).

Furthermore, the CISS MR exhibits fundamentally distinct symmetry\cite{Yang2019,Yang2019b,Naaman2020comment,Yang2020reply,Dalum2019,liu2021chirality} in the current-voltage (I-V) behavior (see Table ~\ref{tab:MR} and Figs. ~\figref{fig:device}{b-c}) from another known chirality-induced phenomenon called electrical magnetochiral anisotropy (EMCA)\cite{Rikken2001}. EMCA refers to the resistance that depends on the current ($\mathbf{I}$) and magnetic field ($\mathbf{B}$) for a chiral conductor, $R^{\chi}=R_0(1+\alpha B^2 + \beta^{\chi} \mathbf{B \cdot I})$, where $\beta^{\chi}=-\beta^{-\chi}$ and
$\chi = \pm$ stands for chirality. The $B^2$ term represents the ordinary MR while the $\mathbf{B \cdot I}$ term represents the unidirectional resistance like a diode (see Fig.\ref{fig:device}b). EMCA was observed in many chiral solids~\cite{Rikken2001,Rikken2019,pop2014electrical,Aoki2019,Inui2020,Shiota2021,Ye2022Te}, where the MR is usually
a few percent or even less. Additionally, the intrinsic magnetic order plays a similar role to the magnetic field and enhances EMCA in experiments\cite{Aoki2019}.
As summarized in Table~\ref{tab:MR}, EMCA respects Onsager's reciprocal relation while CISS MR violates such reciprocity. {Figure~\ref{fig:device}c illustrates the I-V relation of CISS MR.} Onsager's reciprocal theorem originates in the microscopic reversibility of thermodynamic equilibrium and poses strict constraints on macroscopic conductivity\cite{Onsager1931b}. The reciprocity requires that two-terminal conductance remains unchanged as reversing time ($B/M \rightarrow {-B}/{-M}$), i.e., $G(B/M)=G(-{B}/-{M})|_{V\rightarrow 0}$ \cite{Onsager1931b,LANDAU1980,Buttiker1988}, which holds for EMCA and ordinary transport but not for CISS MR. Actually, EMCA can be derived by weakly perturbing the equilibrium ground state to the nonlinear order in the semiclassical theory ~\cite{Ideue2017,Liu2021UMR,kaplan2022unification} while CISS MR will require understanding the far out-of-equilibrium phase. We stress that device ground states for $\pm M$ at thermodynamic equilibrium should be rigidly equivalent because they are time-reversal partners to each other and the equilibrium phase is unchanged in time by definition. Then, it is puzzling to observe varied zero-bias conductance between $\pm M$ in a CISS device.

\begin{table}
\small
\caption{Summary of CISS MR and EMCA in experiments. The conductance is represented by $G(V,M)$, which depends on the bias voltage ($V$) and electrode magnetization ($M$).  Representative values of resistance and magnetoresistance (MR) are taken from Refs.\citeonline{Kiran2016,Naaman2020comment,Kim2021,Qian2022,Al-Bustami2022,Liu2020} for CISS-MR and Refs.\citeonline{Rikken2001,Rikken2019,pop2014electrical,Aoki2019,Inui2020,Shiota2021,Ye2022Te} for EMCA.}
\label{tab:MR}
\begin{tabular}{c|p{2.8cm}p{2.4cm} p{4.1cm}}
\hline
MR-type & Materials & Resistance & Onsager's Relation\\
\hline
CISS-MR & Insulating chiral molecules & $1\sim1000~\mathrm{M\Omega}$  $MR \geq 60\%~~~~$  &  Violated ~~~~~~~~~~~~~~~~~~$G(M)\neq G(-{M})|_{V\rightarrow0}$
\\
\hline
EMCA    & Chiral metals/ semiconductors & $\mathrm{1\sim100 ~\Omega}$ $MR \sim 1~\%~~~~$ or smaller & Preserved ~~~~~~~~~~~~~~~~$G(M)= G(-{M})|_{V\rightarrow0}$\\
\hline
\end{tabular}
\end{table}

% \begin{tabular}{c|p{2.8cm}p{2.4cm} p{4.6cm} p{4.1cm}}
% \hline
% MR-type & Materials & Resistance & Symmetry & Onsager's Relation\\
% \hline
% CISS-MR & Chiral molecules (insulators) & $1\sim1000~\mathrm{M\Omega}$  $MR \geq 60\%~~~~$  &  $G(V,M) \neq G(-{V},-{M})$  
% $G(V,M)>G(V,-{M})$ $G(-{V},M)>G(-{V},-{M})$  &  Violated ~~~~~~~~~~~~~~~~~~$G(M)\neq G(-{M})|_{V\rightarrow0}$
% \\
% \hline
% EMCA    & Metals / doped semiconductors & $\mathrm{1\sim100 ~\Omega}$ $MR \sim 1~\%~~~~$ or smaller & $G(V,M)=G(-{V},-{M})$ $G(V,M)>G(V,-{M})$ $G(-{V},M)<G(-{V},-{M})$   & Preserved ~~~~~~~~~~~~~~~~$G(M)= G(-{M})|_{V\rightarrow0}$\\
% \hline
% \end{tabular}
% \end{table}

%Previous theoretical works\cite{Yang2019b,liu2021chirality} indeed showed that a chiral material presents EMCA-like transport (Fig. \figref{fig:device}{b}) when perturbating the ground state. 
The chiral molecule was commonly regarded as a spin filter~\cite{Naaman2012,Naaman2019} in which electrons exhibit opposite spin polarization after transmitting through the chiral molecule from opposite directions.  
However, recent theoretical works \cite{Yang2019,Yang2019b,Wolf2022} pointed out that reflected electrons by the chiral molecule exhibit the same spin polarization as transmitted ones and the chiral molecule actually acts as a spin polarizer\cite{Wolf2022}. 
As illustrated in Fig. \figref{fig:device}{a}, the right (left) moving electrons can be polarized to spin right (left) for both transmitted and reflected electrons. Such a direction-dependent spin polarization was also revealed by recent \textit{ab initio} calculations~\cite{Naskar2023}. The right(left)-moving electrons generate low (high) resistance if their spin is parallel (anti-parallel) to the polarization of FM electrode. Therefore, the I-V curve exhibits a diode-like feature (see Fig.~\figref{fig:device}{b}), i.e., current-direction-dependent conductance. The sign of reification switches if reversing the magnetization or molecule chirality, presenting EMCA.
The spin polarization in the context of CISS directly leads to the EMCA transport, consistent with recent experiments with chiral conductors ~\cite{Rikken2001,Rikken2019,pop2014electrical,Aoki2019,Inui2020,Shiota2021,Ye2022Te}. 
To understand the CISS MR from chiral molecules, however, we need a mechanism beyond the spin polarization because of the fundamental symmetry reason, regardless of the origin of spin-orbit coupling (SOC) from metal contacts or organic molecule~\cite{Gersten2013,liu2021chirality,adhikari2023interplay}.

In the zero bias limit, many CISS experiments\cite{Kiran2016,Naaman2020comment,Liu2020,Liu2022experiment} showed a clear violation of Onsager's relation by $G(M)\neq G(-{M})|_{V\rightarrow0}$, indicating that switching $M$ drives the system to different {metastable} states and consequently leads to varied conductance. If so, what kind of {metastable} states matter here? Given that a CISS device and EMCA share the same symmetry condition, both inversion symmetry-breaking and time-reversal symmetry-breaking, what causes different transport behaviors between CISS and EMCA? Answers to these questions will help to understand the nature of CISS. In experiments, EMCA involves metals or doped semiconductors, while CISS MR commonly measures insulating chiral molecules \textcolor{black}{(see Table ~\ref{tab:MR})}. In CISS, for example, the typical resistance is in the $\mathrm{G\Omega}$ scale 
\cite{Xie2011}. Therefore, we speculate that the insulating nature of chiral molecules can be essential for understanding CISS MR, which was rarely appreciated in the literature. 
In a molecular tunneling junction, it is natural to use the potential barrier ($U$) to characterize the tunneling conductance. Then we aim to resolve how electrode magnetization ($\pm M$) or chirality modifies the potential barrier when charge flows through the CISS device. Some recent works\cite{Dalum2019,Dubi2021,das2022temperature,hedegaard2023spin} discussed that induced spin accumulation might change the effective potential at non-equilibrium. But this spin accumulation scenario would apply to chiral conductors \textcolor{black}{ and thus, fail to distinguish chiral insulators from chiral metals. }
In this work, we explore the induced charge trapping that directly modulates the potential, which is significant for the insulating chiral molecule but negligible for the chiral metals/doped-semiconductors. It should be noted that carrier trapping is a commonly observed phenomenon in molecular transport \cite{Coropceanu2007,Kaake2010,Haneef2019,sachnik2023elimination} and usually exhibit a long-lived lifetime \cite{Novembre2008memory,Son2009memory,Burgt2018,Park2022}.

In this article, we propose that the CISS MR originates in charge-trapping-induced tunneling-barrier modification (as illustrated in Fig.~\figref{fig:device}{e}).
%, which is indirectly relevant to the spin polarization scenario. 
At the FM-molecule interface (Fig.~\figref{fig:device}{d}) with coexisting magnetism, chirality, and SOC, a non-Hermitian skin effect (NHSE)~\cite{Yao2018,Murakami2019PRL,Zhang2020,YYF2020PRL,OkumaSatoReview,LeeCHReview} appears, which is widely demonstrated in topological systems like photonic lattices and quantum devices ~\cite{Sato2019PRX,XuePeng2020,Thomale2020,Ghatak2020,Fulga2024NP}. NHSE generates extensive exponentially localized eigenstates at two sides of the interface due to dissipation as current flows (see Fig.~\figref{fig:device}{e}). 
The localization direction, which reverses as flipping magnetism or chirality, leads to occupied or empty impurity/defect levels in the molecule side of the interface, and thus generates electron/hole-trapping,\textcolor{black}{ which we call the magnetochiral charge pumping effect.}
The trapped charge has a long lifetime as a metastable state (Fig.~\figref{fig:device}{f}), survives at zero-bias, and consequently alters the electron tunneling barrier in the whole device. 
Therefore, CISS MR refrains from Onsager's relation due to charge trapping and can show significantly larger MR than expected from the electrode spin polarization. The charge-trapping model is consistent with the experimental observation that local charging at the FM-molecule interface lasts long and modifies the surface potential ~\cite{Abendroth2019,Ghosh2020}.  
%Because NHSE is scaled by SOC here, the CISS MR magnitude can be controlled by electrode SOC~\cite{Gersten2013,liu2021chirality} as verified by a recent experiment\cite{Liu2022experiment}.
We further anticipate CISS MR may appear without FM electrodes but with an external magnetic field because NHSE also exists in this case, \textcolor{black}{as verified by very recent experiments~\cite{Wu2024gold,Verhage2024crystal}.}
Because charge trapping requires stabilization by the insulating layer, we predict that CISS MR will diminish and evolve into EMCA if the chiral insulator becomes more metallic. 
The magnetochiral charge pumping  will provide further insights for the magnetism-modulated charge transfer in chirality-related chemical and biological reactions~\cite{Banerjee2020,Ghosh2021,Naaman2022,lu2023beyond,zuo2023mechano}. 

\section{Results}
The induced spin polarization in chiral molecules was frequently regarded as the MR ratio in literature. However, resistance rather than spin is the directly measured quantity in transport. Thus, we circumvent the illusive spin polarization~\cite{Liu2023spin} and focus on MR in this work. 
As discussed above, the nonequilibrium spin polarization in the context of CISS directly leads to EMCA, which is independent of model details such as the origin of SOC. 
In the following, we will reveal NHSE at the FM-molecule interface. Because NHSE can alter the occupation of defect/disorder states in the molecule side and then lead to charge trapping there.
The trapped charge can survive even at zero bias and modulate the tunneling barrier sensitively, circumventing the Onsager's reciprocal relation. Finally, we will propose a simple tunneling barrier model that can reproduce essential features of CISS MR and extract the barrier information from experiments.

\begin{figure}[p]
    \centering
    \includegraphics[width=1\linewidth]{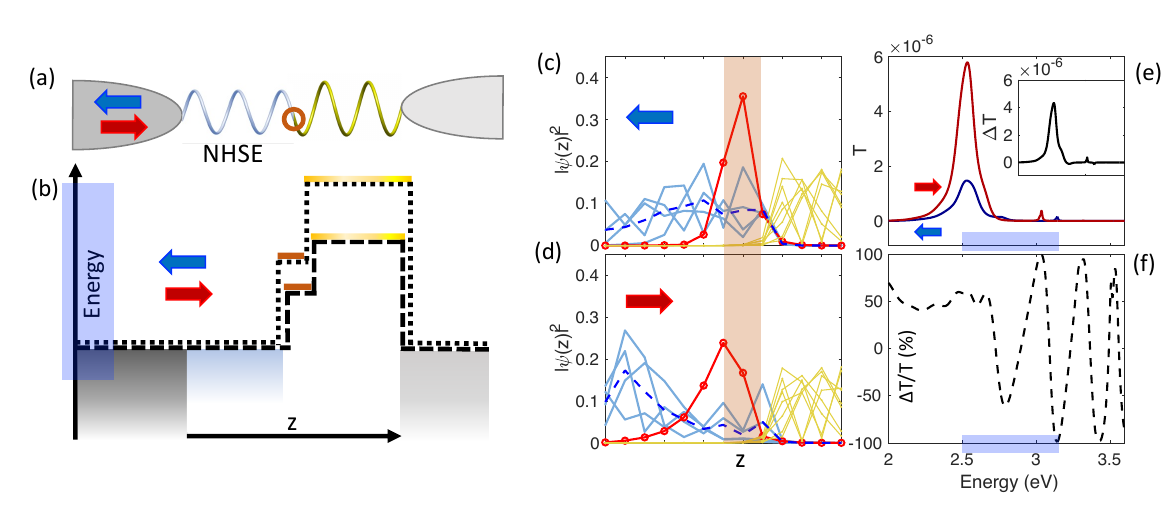}
    \caption{
    \textbf{Barrier modification in a molecular device.} (a) The two-terminal device includes an interface region (with $M$, SOC, and $\chi$) with non-Hermitian skin effect (NHSE) and an ordinary molecule region (with only $\chi$).
    (b) Illustration of the potential profile across the device. The charge trapping sensitively modifies the energy of impurity state (dark orange) and the molecule state (yellow) but leaves the interface region nearly unchanged, because the interface is much more metallic than the molecule region. 
    (c) and (d) show the wavefunction distribution of interface (light blue), impurity (red) and molecule (yellow) states due to NHSE at opposite $M$. The blue dashed curve represents the average of interface wave functions. These states locate in an energy window indicated by the blue shadow along the energy axis in (b), (e) and (f). The impurity occupation probability is evaluated in the red shadow region that is centered at the impurity site. 
    (e) The transmission probability (T) at different chemical potential for opposite $M$ calculated for the two-terminal device shown in (a). (f) shows corresponding ratio of transmission change. 
    }
    \label{fig:NHSE}
\end{figure}

\subsection{Non-Hermitian skin effect}
EMCA can also be understood by NHSE besides the spin polarization scenario. NHSE is widely studied in dissipative (nonequilibrium) systems such as photonic lattices and characterized by exponentially localized eigen wavefunctions at system boundaries, as a topological phenomenon. In realistic materials, NHSE requires dissipation and breaking both time-reversal symmetry and inversion symmetry~\cite{Sato2019PRX,Kawabata2020PRB,YYF2020PRL}. In the CISS device (Fig.~\figref{fig:device}{d} or Fig.~\figref{fig:NHSE}{a}), the FM-molecule interface is the only region combining three ingredients, $M$ (time-reversal breaking), $\chi$ (inversion-breaking), and SOC (coupling $M$ and $\chi$ together), satisfying the NHSE condition. 
In the following, {
we employ a chiral chain model consisting of two parts, as illustrated in Fig.~\figref{fig:NHSE}{a}. The left part represents the interface region, which includes both $M$ and SOC. The right part models the insulating molecule region. Additionally, we introduce an impurity site between these two parts to serve as a charge trapping center. Neither the molecule region or impurity site includes $M$ or SOC, to represent a realistic molecular device.}

We construct a tight-binding model for the chiral chain, incorporating orbital-dependent dissipation in the left part (see SI for details). This left part exhibits NHSE, characterized by localized eigenstates at the boundaries. As illustrated in Fig.~\figref{fig:NHSE}{c-d}, positive magnetization ($+M$, $\Rightarrow$) causes wave functions to localize at the left open boundary, while negative magnetization ($-M$, $\Leftarrow$) generates states localized in the junction region. The impurity state, due to its coupling with the interface part, is highly sensitive to NHSE. Depending on the magnetization ($\pm M$), its wave function is either drawn away from or more localized at the impurity site (see more details in SI). In contrast, wave functions in the molecule part are only marginally affected by NHSE. NHSE induces unidirectional tunneling through the system. For instance, the left-side localization of eigenstates (Fig.\ref{fig:NHSE}d) enhances electron tunneling from right to left while suppressing tunneling in the opposite direction, as reported in the literature\cite{YYF2020PRL}. Consequently, the unidirectional tunneling resistance can be reversed by altering either $M$ or $\chi$, giving rise to EMCA which respects the general reciprocity (also see Supplementary Fig. S7). Thus, EMCA serves as the transport signature of NHSE, while NHSE represents the spectroscopic manifestation of the system out of equilibrium. \textcolor{black}{Here, solely NSHE cannot violate the Onsager's relation to produce CISS-type MR.}

We note that NHSE comes from interplay between $M$, $\chi$, SOC, and dissipation. It vanishes if one of them disappears in our model. Specially, dissipation is essential for NHSE. In a CISS device, it originates from the current-related dissipation due to scattering by lattice, disorder, interface, etc. Because the chiral molecule is typically anisotropic, conduction channels with distinct orbital features (e.g., $\pi$ and $\sigma$ orbitals) may dissipate differently. Thus, we include orbital-dependent dissipation in the model. {In addition, the role of dissipation/dephasing was discussed to study the spin polarization in earlier theoretical works on CISS~\cite{Guo2012,Guo2014a,Matityahu2016}.
} We should also note that NHSE cannot emerge from the finite-temperature effect at thermodynamic equilibrium, because energy dissipation and gain between electrons and environment (e.g., lattice vibrations \cite{das2022temperature}) compensate each other in this case. 

\subsection{Charge trapping and barrier modification}
When dissipation diminishes at zero bias, NHSE disappears. Then, merely NHSE cannot explain the violation of Onsager's reciprocity. Therefore, we need the charge trapping effect which survives at zero bias. 
 
Conceive a localized impurity level between left and right parts in the model of Fig.~\figref{fig:NHSE}{a}, which may come from interface states (e.g., thiol bonds \cite{Piccinin2003,Souza2014}), structural defects or extrinsic impurities (e.g. oxygen\cite{Zhuo2009oxygen} or water\cite{Nikolka2019water,Kotadiya2019water}). This state acts as charge(electron/hole)-trapping centers in the insulating molecule side. 
% Carrier trapping is a commonly observed phenomenon in molecular charge transport \cite{Coropceanu2007,Kaake2010,Haneef2019,sachnik2023elimination} and usually exhibit a long-lived lifetime \cite{Novembre2008memory,Son2009memory,Burgt2018,Park2022}. 
The key question is how the charge-trapping can be controlled by $M$ and $\chi$, to which we attribute NHSE. 

As shown above, the impurity state wavefunction is also modified by NHSE. To avoid confusion, we initialize zero spin polarization for the impurity in our model. For simple, we assume that one impurity traps only an electron (the unipolar transport case), which can be easily generalized to the hole trapping or bipolar transport. 
For $-M$ in Fig.~\figref{fig:NHSE}{c}, NHSE tends to keep the impurity state occupied by an electron. Even if it is empty, NHSE enhances electron transfer from left to right to fill the impurity level. In contrast, NHSE keeps the impurity level empty at $+M$. We should stress that the occupied and unoccupied states are metastable on the energy surface (see Fig.~\figref{fig:device}{f}) and remain robust at zero bias. Here, NHSE plays a driving force to swap the impurity level between occupied and empty phases, which depends on $\pm M$ (also $\chi$). The charge trapping can sensitively tune the tunneling barrier in a CISS device, leading to magnetization-(also chirality-) dependent resistance even at zero bias.

{To quantitatively demonstrate the barrier modification, we calculated the zero-bias conductance for the device shown in Fig.\figref{fig:NHSE}{a}. For a given magnetization $M$, we evaluated the charge-trapping probability ($\rho_M$) from the wave function localization at the impurity site (integrated in the red shadow region in Figs.\figref{fig:NHSE}{c-d}, where $\rho_{+M/-M}\approx 0.85/0.45$). When an electron is trapped, it can elevate the electrostatic potential by $V_0$ in the impurity and molecule regions while leaving the interface region largely unchanged, due to the insulating nature of the former and the metallic character of the latter.
We estimated $V_0 = \frac{e^2}{4\pi \epsilon d}\approx 0.4$ eV, considering a $3\sim5$ nm characteristic thickness of the chiral molecule layer and $\epsilon=2.1 \epsilon_0$~\cite{Akkerman2007}. For simplicity, we applied a rigid onsite potential shift of $V_0 \rho_M$ to  impurity and molecule parts. We then calculated the two-terminal tunneling probability for $\pm M$ ($T_{\pm M}$) after attaching a ferromagnetic electrode and an ordinary lead. To demonstrate the violation of Onsager's relation, we focus on the zero-bias conductance at a given chemical potential (see Fig.~\figref{fig:NHSE}{e}), in which the dissipation vanishes but charge trapping remains. The resulting $T\ll 1$ because it originates from potential barrier tunneling through the insulating molecule. The electron barrier height equates to the energy difference between the electrode's chemical potential and the molecule's lowest unoccupied molecular orbital(see Fig.~\figref{fig:NHSE}{b}). Consequently, even a small barrier modification due to different $\rho_M$ values can lead to a large change in $T$ ($\Delta T = T_{-M}-T_{+M}$), characterizing the violation of Onsager's relation. 
Figure ~\figref{fig:NHSE}{f} shows that the ratio of $T$ change [($\Delta T / (T_{-M}+T_{+M})$)] can exceed 50\% and even approach 100\% for certain chemical potentials. This further indicates that both the sign and magnitude of CISS MR depend not only on $\chi$ and $M$ but also on device-specific details such as the chemical potential.
Additionally, we note that more complex potential profiles also yield large $T$ change ratio (see SI) because $T$ is quite sensitive to potential changes in the tunneling regime. }

We stress that the spin polarization may be relevant to CISS MR but cannot explain the reciprocity violation. The trapped charge by the insulating molecule is a key ingredient to break the microscopic reversibility and circumvent the Onsager relation. This also explains why conducting chiral crystals, 
\textcolor{black}{
in which charge cannot be trapped or trapped charge has marginal influence in resistance,
} exhibit EMCA rather than CISS MR.  
{Our model also rationalizes why the CISS MR can  significantly exceed the spin polarization ratio in the FM electrode. This is because the modified barrier dramatically amplifies the tunneling conductance, as demonstrated by our calculations. } In addition, the FM-molecule interface model naturally breaks the 180$^\circ$ rotation symmetry of the device, coinciding with a recent discussion on the symmetry requirement of CISS MR~\cite{rikken2023comparing}.

\subsection{An effective barrier model.}
The trapped charge (electron or hole) can modify the injection barrier at the metal-molecule interface and/or change the barrier profile (e.g., shape, height, width) in a complicated way. For further simplicity, we present an effective model by considering  a rectangular barrier with height $U_0$ and a barrier modulation $\delta U$, in spirit of the widely adopted Simmons model to simulate the electron transport in molecular junctions\cite{Simmons1963}. If the molecule conducts electrons, electron or hole trapping indicates an increase or decrease in the barrier, respectively, and vice versa for the hole-conducting case.

\begin{figure}
    \centering
    \includegraphics[width=0.6\linewidth]{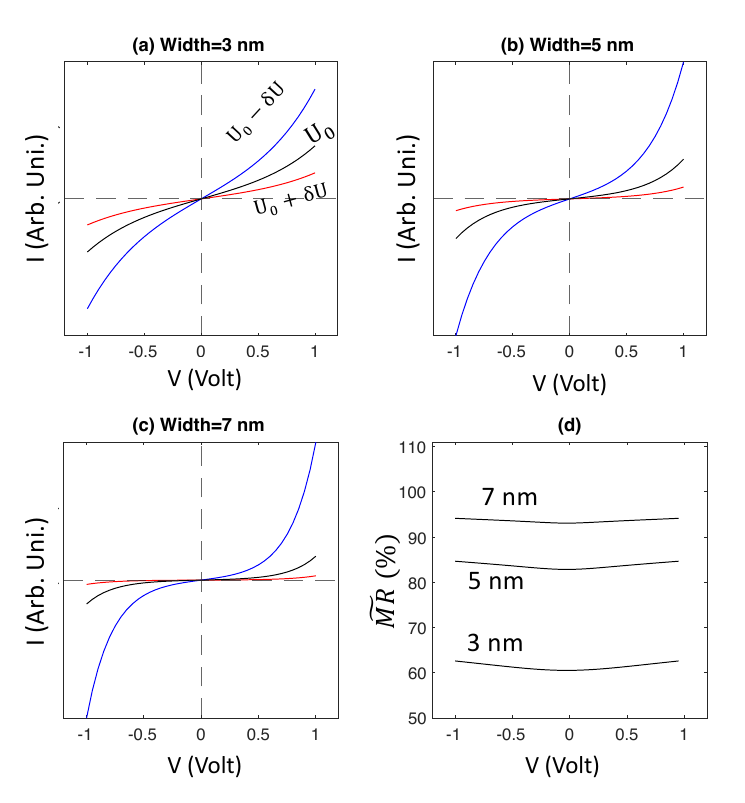}
    \caption{\textbf{The I-V curves calculated from Eq.~\eqref{Eq:tunneling}.} (a)-(c) The MR increases dramatically as increasing width. (d) The normalized MR [$\widetilde{MR}$, defined in Eq.~\eqref{Eq:MR}] increases with the width.    
    }
    \label{fig:MR}
\end{figure}

%If we consider experimental parameters from a prototype molecule device\cite{Akkerman2007} with a barrier height $U_0 \sim 5$ eV and dielectric constant $\epsilon = 2.1 \epsilon_0$ where $\epsilon_0$ is the vacuum permittivity, the induced barrier change is about $\delta U \sim \pm {e \delta q}/(4\pi \epsilon d) = \mp 0.2 $ eV for the typical molecule length $d = 5$ nm. 
We can estimate the transport through a rectangular barrier $U=U_0+\delta U$ in the case of $eV < U$,
\begin{equation}
    I=g [(U+eV/2)e^{-A*\sqrt{U+eV/2}}-(U-eV/2)e^{-A*\sqrt{U-eV/2}}], \label{Eq:tunneling}
\end{equation}
where $A=\frac{4\pi l \sqrt{2m^*}}{h}$, $l$ is the barrier width, $g$ is a constant in unit of conductance, $m^*$ is the effective mass. Here, we symmetrize Eq.~\ref{Eq:tunneling} for $\pm V$ compared to the Simmons model~\cite{Simmons1963}.

As shown in Fig.~\ref{fig:MR}, I-V curves calculated by Eq.~\ref{Eq:tunneling} well reproduce the symmetry of CISS-MR and the variation of zero-bias conductance. We take $U_0=5~eV, \delta U =0.2 ~eV$ and $m^* = 0.28 ~m_e$ \cite{Akkerman2007} where $m_e$ is the free electron mass. The MR ratio enlarges slightly as increasing bias but increases dramatically as increasing the barrier width. 
In literature \cite{Naaman2015}, the CISS driven spin polarization ($P$) is frequently defined via the current change for $\pm M$,
\begin{equation}
P=\frac{|I(+M)-I(-M)|}{I(+M)+I(-M)},    
\end{equation}
which leads to an unphysical abrupt drop of $P$ near the zero bias. 
Instead, we define a similar quantity termed normalized MR ($\widetilde{MR}$) by the change of conductance $G=dI/dV$ (or resistance) as,
\begin{equation} \label{Eq:MR}
    \widetilde{MR}=\frac{|G(+M)-G(-M)|}{G(+M)+G(-M)} \equiv \frac{|R(-M)-R(+M)|}{R(-M)+R(+M)},
\end{equation}
which exhibits a more reasonable bias dependence near the zero bias region. 
Consistent with experimental observations~\cite{Kulkarni2020,Mishra2020,Al-Bustami2022}, we find that $\widetilde{MR}$ increases as increasing the barrier width because barrier width is proportional to $A$ in Eq.~\eqref{Eq:tunneling}. 
As shown in Fig.~\ref{fig:MR}, one can reach a high $\widetilde{MR}$ up to 100\% by engineering  device parameters such as the barrier width. 

{We stress that $\widetilde{MR}$ qualitatively  
differs from the so-called spin polarization ($P_{\chi}$) in CISS literature despite that they are frequently considered to be equivalent to each. }
If we follow the tunneling magnetoresistance picture~\cite{Julliere1975TMR} regarding tunnelling between the FM electrode with spin polarization $P_{FM}$ and chiral molecule, we get $\widetilde{MR}=P_{\chi}P_{FM} \neq P_{\chi}$ except $P_{FM} = 100\%$ which is unrealistic.
Known the low $P_{FM}$ in ordinary FM, $\widetilde{MR} \le P_{FM}$ obviously contradicts the high $\widetilde{MR}$ in experiments. Thus, we suggest that $\widetilde{MR}$ may serve a more proper terminology to characterize CISS transport compared to the so-called spin polarization.
{In our theory, the large $\widetilde{MR}$ value is caused by the nonlinear amplification of the barrier tunneling, which is indirectly related to spin polarization.}

\section{Discussions and Outlook}

\begin{figure}
    \centering
    \includegraphics[width=1.05\linewidth]{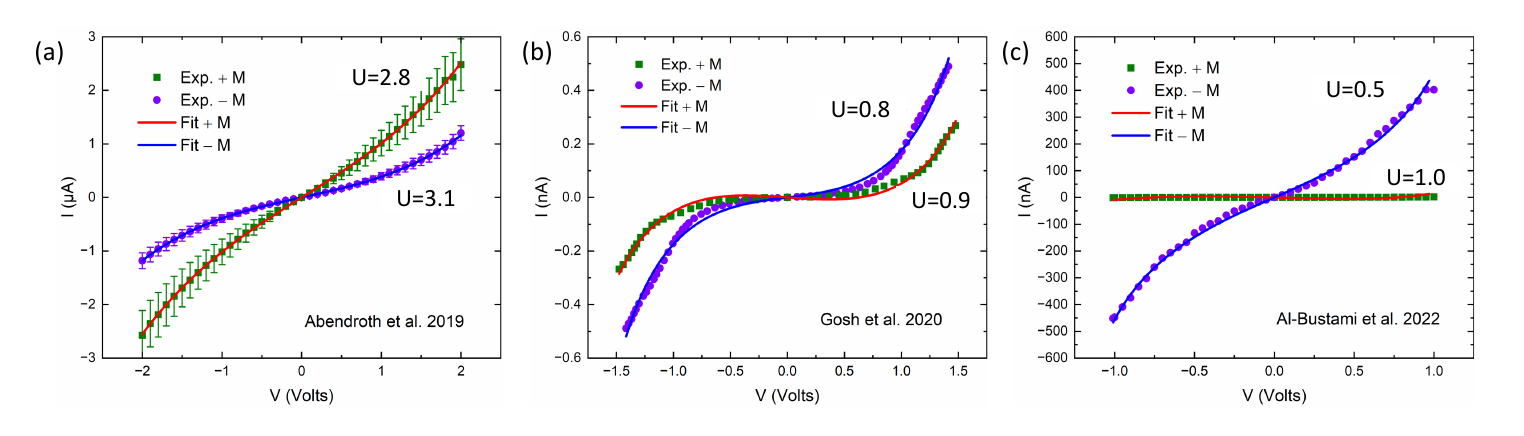}
   \caption{\textbf{Fitting the experimental results with the barrier model.} The experimental data were obtained from (a) Ref.~\citeonline{Abendroth2019}, (b) Ref.~\citeonline{Ghosh2020} and (c) Ref.~\citeonline{Al-Bustami2022}. The extracted barrier height (U) is marked with corresponding I-V curve. 
    }
    \label{fig:fitting}
\end{figure}

\begin{figure}
    \centering
    \includegraphics[width=0.6\linewidth]{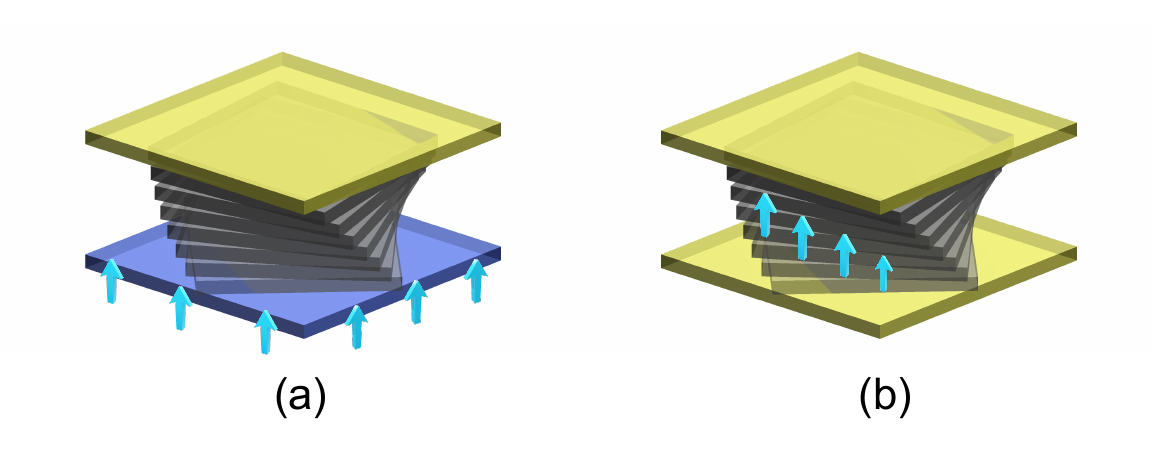}
   \caption{\textbf{CISS platforms by
    twisted van der Waals layers sandwiched between vertical electrodes.}  The electrodes can be either metals or metallic/semimetallic van der Waals materials. Blue arrows indicate the magnetic material. Magnetism can appear in one/both electrodes (a) and/or (b) in the twisted material. If the twisted layers are insulating, CISS-type MR can occur. If it is more conducting, it may exhibit EMCA. 
    The role of SOC can be examined in different electrodes like graphite and transition metal dichalcogenides. 
    }
    \label{fig:twisted}
\end{figure}

The charge trapping due to NHSE circumvents Onsager's relation and reproduces the major symmetry of CISS transport. The charge trapping may occur at the molecule-metal interface, defects, disorders, or impurities, and its microscopic origin may vary among different molecules and devices. 
Nevertheless, we can make insightful predictions from the proposed mechanism that relies on NHSE and charge trapping.

(i) If a device traps both electrons and holes, one can expect the ideal barrier modification as $U=U_0 \pm \delta U$ for $\pm M$. (ii) If a device traps only electrons/holes, the barrier is switched between $U_0$ and $U=U_0 \pm \delta U$. Here, a salient feature is that the conductance for one of $\pm M$ is comparable to the $M=0$ case, {showing asymmetric MR between $\pm M$}. 
(iii) If the device does not efficiently trap charge, the CISS MR may diminish and change to EMCA. This is consistent with the EMCA observed in metallic or nearly metallic chiral solids \cite{Rikken2001,Rikken2019,pop2014electrical,Aoki2019,Inui2020,Shiota2021} where charge trapping can be screened by the metallic background. EMCA may also occur for an insulating chiral molecule if the device is ideally clean without a charge-trapping center. (iv) Because NHSE and EMCA does not necessarily require an FM electrode, we anticipate that CISS MR may also occur with ordinary electrodes in an external magnetic field if the charge trapping condition is satisfied. {For example, electrons in a metallic chiral crystal can feel the magnetic field by the Lorentz force, and charge trapping may occur at the crystal interface with another insulating layer that is essential to determine the sign of MR. \textcolor{black}{
In this case, SOC in materials/electrodes is not essential for CISS MR because magnetic field directly couples to the electron motion (orbital).}
}
(v) The amplitude of EMCA (also NHSE) is scaled by SOC in electrodes\cite{liu2021chirality,Gersten2013} because organics molecules usually have negligible SOC. So CISS MR is also scaled by electrode SOC. The electrode dependence was confirmed by observing significantly different MR with Au and Al electrodes in Ref.~\citeonline{Liu2022experiment} recently. 
{
We should stress that CISS-type MR is directly determined by the charge trapping effect despite that it is related to the SOC. In some cases, it is possible to observe that CISS MR is less sensitive to SOC while coexisting EMCA is more sensitive to SOC. For example, when SOC is strong enough to enforce charge trapping probability nearly 100\%, further increasing SOC will hardly change CISS MR despite of enhancing EMCA.}
(vi) The charge trapping usually comes with local structural distortion at the interface or inside the molecule. Such distortion may be detected via the change of atomic bonding, for example, by the Raman spectrum.

Furthermore, we can include the I-V asymmetry  due to EMCA or device asymmetry with 
~\eqref{Eq:tunneling} and provide a general formula to describe the I-V relation for chiral molecular devices, 
\begin{equation}
    I = g [(U+eV/2)e^{-A*\sqrt{U+eV/2}}-(U-eV/2)e^{-A*\sqrt{U-eV/2}}] (1 - \gamma V) , \label{Eq:IVall}
\end{equation}
where $\gamma = (\gamma_0 + \gamma^{\chi}_M)/R_0$ includes $\gamma^{\chi}_M$ due to EMCA and $\gamma_0$ that is a constant caused by device asymmetry such as different electrodes or molecular dipole. Equation \eqref{Eq:IVall} indicates a crossover from  CISS MR to EMCA if $\gamma^{\chi}_M$ is dominant or potential change ($\delta U$) is negligible, as discussed above. 

We suggest that one can extract key parameters, including $\gamma$ and $U$,  by fitting experiments with Eq.~\eqref{Eq:IVall}.  For example, we successfully fit the original experimental I-V relation using Eq.~\eqref{Eq:IVall} in Fig.~\ref{fig:fitting}. 
For the same device with a given chirality, we globally fit two sets of I-V curves with $\pm M$ by using the same $g$ and $A$ because they are invariant for $\pm M$.(see more details in SI) We obtain the barrier change $\delta U =$ 0.3, 0.1 and 0.5 eV for experiments by Abendroth \etal\cite{Abendroth2019},  Gosh \etal\cite{Ghosh2020} and  Al-Bustami \etal\cite{Al-Bustami2022}. The large $\delta U $ (also large $\delta U / U$) well rationalizes that Al-Bustami \etal observed a much larger $\widetilde{MR}$ than the other two works.
Even if we know little atomic configuration of the device, we can still use the barrier model to rationalize CISS experiments.

Furthermore, Abendroth $et~al.$ \cite{Abendroth2019} and Gosh $et~al.$\cite{Ghosh2020} also observed the work function change $\sim 0.1$ eV due to switching magnetization for chiral films on the FM substrate
and discussed its correlation to CISS. As shown above, we obtain a similar magnitude in the charge tunneling barrier change ($\delta U$) by fitting their transport results. The barrier is related to the energy offset between metal Fermi surface and molecule levels, and work function is related to the difference between metal Fermi surface and vacuum. Because both can be modified by the surface/interface charging effect, it is not surprising to expect the comparable magnitude in changes due to magnetization flip. 
In addition, Abendroth \etal showed negligible surface potential change after they neutralized the surface charging, indicating a direct correlation between the charging and CISS~\cite{Abendroth2019}.
This further supports that the trapped-charge-induced barrier modulation can be a key ingredient to understand CISS transport, which correlates to the work function change. 

\textcolor{black}{
The magnetochiral charge pumping is a dynamical process characterized by complex interfacial interactions. When current flows through a chiral molecular device, magnetization/magnetic field switching first reverses the interface wavefunction localization direction via NHSE.
This process requires a finite time to modify the charge trapping state, which is typically close to or larger than the structural relaxation time due to associated lattice distortions. The barrier change (0.1$\sim$1 eV) cannot be compensated by the Zeeman energy from electrode magnetization flipping (around 1 meV).  In this open system, the required energy to alter the potential profile is provided by the external current flow. Future work should focus on establishing a quantitative description of the dynamical interplay between NHSE and charge trapping to elucidate the underlying microscopic mechanisms.}

{Our work reveals a fundamental mechanism how magnetism or magnetic field modifies the charge distribution at the metal-organic interface through nonequilibrium dynamics. Besides chiral molecular transport, our work can also explain the magnetic substrate-modulated charge-reorganization in protein-protein association~\cite{Banerjee2020,Ghosh2021},
\textcolor{black}{
in which the magnetochiral charge pumping changes the electrostatic potential of proteins by switching the substrate magnetization. }
It paves a pathway to modulate chemistry and biology processes, which commonly involve chiral materials, by controlling the charge transfer or the long-range electrostatic potential.  
}

We have another outlook that the recently discovered twisted van der Waals materials\cite{Bistritzer2011,Cao2018Mott,Cao2018SC} provide versatile platform for the CISS effect.  
In twisted bilayer graphene and twisted transition-metal dichalcogenides\cite{Wu2019tmd,Zhang2020tmd,Ghiotto2021tmd,Shan.Xu.2022}, the in-plane transport has been extensively studied for flat-band-driven phenomena such as the unconventional superconductivity, Mott insulating states, and Chern insulators. We conceive a two-terminal vertical device where two electrodes sandwich twisted bilayers or multiple layers, as illustrated in Fig.~\ref{fig:twisted}. It was recently proposed that flat bands significantly enhance the EMCA effect in twisted bilayer graphene\cite{Liu2021UMR}. The EMCA comes from the coexistence of magnetization (or magnetic field) and chirality. Since twisted layers are chiral, we introduce magnetism to either one electrode and/or twisted layers.
Many van der Waals materials are magnetic, such as insulating CrI$_3$\cite{Wang2020}, semiconducting MnBi$_2$Te$_4$\cite{Lian2020}, and metallic Fe$_3$GeTe$_2$\cite{kim2020observation} and Cr$_3$Te$_4$ \cite{Qian2022}.
Additionally, twisting between van der Waals electrodes can also play the role of chirality. The vertical resistance \cite{Inbar2023} can be measured for the opposite magnetization and for opposite twist angles. If chiral layers are insulating, we anticipate CISS-like MR. If they are metallic, we expect EMCA-like MR. In most of the present CISS devices, a thin layer of gold is technically required to protect the ferromagnetic film. This gold layer results in difficulties in examining the influence of substrate SOC\cite{liu2021chirality,Gersten2013}. {In Fig. \ref{fig:twisted}, electrodes can also be made of van der Waals materials (such as graphite) with negligible SOC, which circumvents this problem as well as the complexity of metal-organic molecule junction.} Therefore, van der Waals materials provide a versatile platform to investigate the relation of CISS with SOC, magnetism, and the chiral layer 
insulation. It would be more exciting to examine the influence of flat bands\cite{Liu2021UMR,Bistritzer2011}, correlation effects, and even superconductivity in CISS \cite{Alpern2016}.

{
Note: During the review of the manuscript, our predictions (ii)\&(iv) on the asymmetric magnetoresistance in an external magnetic field were realized in chiral gold nanocrystals~\cite{Wu2024gold}, in which metallic gold core shows the chiral morphology and insulating surface capping layer traps the charge in the magnetochiral charge pumping. 
\textcolor{black}{A similar asymmetric magnetoresistance was also observed in a chiral supermolecular crystal with solely a magnetic field where the charge trapping effect was observed~\cite{Verhage2024crystal}. }
Our prediction (v) on the coexistence of EMCA and CISS was reported in chiral single-molecule junction measurements~\cite{singh2024single} very recently, 
\textcolor{black}{
in which EMCA is sensitive to SOC while CISS is nearly independent of SOC, and the reciprocal relation is violated at zero bias.}
}

\section*{Data availability}

The data for the charge barrier model in a chiral molecular and the transmission probability for the two-terminal device is available in Zenodo at \url{https://doi.org/10.5281/zenodo.14205821}. Other Supplementary information that support this work are available upon request to the corresponding author. 

\section*{Code availability}

The Mathematica code used to calculate the charge barrier model, modify the impurity potential under reversed magnetism, and to calculate the transmission probability using the non-equilibrium Green's function formalism is available in Zenodo at \url{https://doi.org/10.5281/zenodo.14205821}. 

\section*{Acknowledgements}
We thank Yossi Paltiel, John M. Abendroth, Tianhan Liu, Peng Xiong, and Paul Weiss for sharing their experimental data and helpful discussions. We are grateful for the inspiring discussions with Yuval Oreg, Eric Akkermans, Ron Naaman, Yizhou Liu, Tobias Holder, ChiYung Yam, Per Hedegard, Oren Tal, D. H. Waldeck, David Mukamel, Oren Raz, Xi Dai, Dali Sun and Zhenfei Liu. 
B.Y. acknowledges the financial support by the MINERVA Stiftung, the European Research Council (ERC Consolidator Grant ``NonlinearTopo'', No. 815869), and the Israel Science Foundation (ISF, No. 2932/21). K.S. and K.Z. are supported in part by the Office of Naval Research (MURI N00014-20-1-2479). 

\section*{Author Contributions Statement}
B.Y. conceived the project and wrote the manuscript with inputs from all authors. Y.Z. and J.X. performed barrier tunneling calculations. K.Z. and K.S. performed non-Hermitian skin effect and two-terminal transmission calculations. All authors analyzed the results. 

\section*{Competing Interests Statement}
The authors declare no competing interests.

\section*{}

\newpage
\section*{References}
%\bibliography{references-chiral}

\begin{thebibliography}{10}
\expandafter\ifx\csname url\endcsname\relax
  \def\url#1{\texttt{#1}}\fi
\expandafter\ifx\csname urlprefix\endcsname\relax\def\urlprefix{URL }\fi
\providecommand{\bibinfo}[2]{#2}
\providecommand{\eprint}[2][]{\url{#2}}

\bibitem{kelvin1894molecular}
\bibinfo{author}{Kelvin, W. T.~B.}
\newblock \emph{\bibinfo{title}{The molecular tactics of a crystal}}
  (\bibinfo{publisher}{Clarendon Press}, \bibinfo{year}{1894}).
\newblock
  \urlprefix\url{https://archive.org/details/moleculartactics00kelviala/}.

\bibitem{Gohler2011}
\bibinfo{author}{Gohler, B.} \emph{et~al.}
\newblock \bibinfo{title}{{Spin Selectivity in Electron Transmission Through
  Self-Assembled Monolayers of Double-Stranded DNA}}.
\newblock \emph{\bibinfo{journal}{Science}} \textbf{\bibinfo{volume}{331}},
  \bibinfo{pages}{894 -- 897} (\bibinfo{year}{2011}).
\newblock
  \urlprefix\url{https://www.science.org/doi/full/10.1126/science.1199339}.

\bibitem{Naaman2012}
\bibinfo{author}{Naaman, R.} \& \bibinfo{author}{Waldeck, D.~H.}
\newblock \bibinfo{title}{{Chiral-Induced Spin Selectivity Effect}}.
\newblock \emph{\bibinfo{journal}{The Journal of Physical Chemistry Letters}}
  \textbf{\bibinfo{volume}{3}}, \bibinfo{pages}{2178--2187}
  (\bibinfo{year}{2012}).
\newblock \urlprefix\url{https://pubs.acs.org/doi/10.1021/jz300793y}.

\bibitem{Naaman2019}
\bibinfo{author}{Naaman, R.}, \bibinfo{author}{Paltiel, Y.} \&
  \bibinfo{author}{Waldeck, D.~H.}
\newblock \bibinfo{title}{{Chiral molecules and the electron spin}}.
\newblock \emph{\bibinfo{journal}{Nature Reviews Chemistry}}
  \textbf{\bibinfo{volume}{3}}, \bibinfo{pages}{250--260}
  (\bibinfo{year}{2019}).
\newblock \urlprefix\url{https://www.nature.com/articles/s41570-019-0087-1}.

\bibitem{evers2021theory}
\bibinfo{author}{Evers, F.} \emph{et~al.}
\newblock \bibinfo{title}{Theory of chirality induced spin selectivity:
  Progress and challenges}.
\newblock \emph{\bibinfo{journal}{Advanced Materials}}
  \textbf{\bibinfo{volume}{34}}, \bibinfo{pages}{2106629}
  (\bibinfo{year}{2022}).
\newblock
  \urlprefix\url{https://onlinelibrary.wiley.com/doi/full/10.1002/adma.202106629}.

\bibitem{yan2024structural}
\bibinfo{author}{Yan, B.}
\newblock \bibinfo{title}{Structural chirality and electronic chirality in
  quantum materials}.
\newblock \emph{\bibinfo{journal}{Annual Review of Materials Research}}
  \textbf{\bibinfo{volume}{54}}, \bibinfo{pages}{97--115}
  (\bibinfo{year}{2024}).
\newblock \urlprefix\url{https://doi.org/10.1146/annurev-matsci-080222-033548}.

\bibitem{Gersten2013}
\bibinfo{author}{Gersten, J.}, \bibinfo{author}{Kaasbjerg, K.} \&
  \bibinfo{author}{Nitzan, A.}
\newblock \bibinfo{title}{{Induced spin filtering in electron transmission
  through chiral molecular layers adsorbed on metals with strong spin-orbit
  coupling}}.
\newblock \emph{\bibinfo{journal}{The Journal of Chemical Physics}}
  \textbf{\bibinfo{volume}{139}}, \bibinfo{pages}{114111}
  (\bibinfo{year}{2013}).
\newblock \urlprefix\url{https://aip.scitation.org/doi/10.1063/1.4820907}.
\newblock \eprint{1306.4904}.

\bibitem{liu2021chirality}
\bibinfo{author}{Liu, Y.}, \bibinfo{author}{Xiao, J.}, \bibinfo{author}{Koo,
  J.} \& \bibinfo{author}{Yan, B.}
\newblock \bibinfo{title}{Chirality-driven topological electronic structure of
  dna-like materials}.
\newblock \emph{\bibinfo{journal}{Nature Materials}}
  \textbf{\bibinfo{volume}{6}}, \bibinfo{pages}{638–644}
  (\bibinfo{year}{2021}).
\newblock \urlprefix\url{https://www.nature.com/articles/s41563-021-00924-5}.

\bibitem{das2022temperature}
\bibinfo{author}{Das, T.~K.}, \bibinfo{author}{Tassinari, F.},
  \bibinfo{author}{Naaman, R.} \& \bibinfo{author}{Fransson, J.}
\newblock \bibinfo{title}{Temperature-dependent chiral-induced spin selectivity
  effect: Experiments and theory}.
\newblock \emph{\bibinfo{journal}{The Journal of Physical Chemistry C}}
  \textbf{\bibinfo{volume}{126}}, \bibinfo{pages}{3257--3264}
  (\bibinfo{year}{2022}).
\newblock \urlprefix\url{https://pubs.acs.org/doi/10.1021/acs.jpcc.1c10550}.

\bibitem{hedegaard2023spin}
\bibinfo{author}{Hedeg{\aa}rd, P.}
\newblock \bibinfo{title}{Spin dynamics and chirality induced spin
  selectivity}.
\newblock \emph{\bibinfo{journal}{The Journal of Chemical Physics}}
  \textbf{\bibinfo{volume}{159}} (\bibinfo{year}{2023}).
\newblock
  \urlprefix\url{https://pubs.aip.org/aip/jcp/article/159/10/104104/2910483}.

\bibitem{Xie2011}
\bibinfo{author}{Xie, Z.} \emph{et~al.}
\newblock \bibinfo{title}{{Spin specific electron conduction through DNA
  oligomers.}}
\newblock \emph{\bibinfo{journal}{Nano letters}} \textbf{\bibinfo{volume}{11}},
  \bibinfo{pages}{4652--5} (\bibinfo{year}{2011}).
\newblock \urlprefix\url{https://pubs.acs.org/doi/10.1021/nl2021637}.

\bibitem{Kiran2016}
\bibinfo{author}{Kiran, V.} \emph{et~al.}
\newblock \bibinfo{title}{{Helicenes-A New Class of Organic Spin Filter}}.
\newblock \emph{\bibinfo{journal}{Advanced Materials}}
  \textbf{\bibinfo{volume}{28}}, \bibinfo{pages}{1957--1962}
  (\bibinfo{year}{2016}).
\newblock
  \urlprefix\url{https://onlinelibrary.wiley.com/doi/full/10.1002/adma.201504725}.

\bibitem{Varade2018}
\bibinfo{author}{Varade, V.} \emph{et~al.}
\newblock \bibinfo{title}{{Bacteriorhodopsin based non-magnetic spin filters
  for biomolecular spintronics}}.
\newblock \emph{\bibinfo{journal}{Physical Chemistry Chemical Physics}}
  \textbf{\bibinfo{volume}{20}}, \bibinfo{pages}{1091--1097}
  (\bibinfo{year}{2018}).
\newblock
  \urlprefix\url{https://pubs.rsc.org/en/content/articlelanding/2018/cp/c7cp06771b}.

\bibitem{Liu2020}
\bibinfo{author}{Liu, T.} \emph{et~al.}
\newblock \bibinfo{title}{Linear and nonlinear two-terminal spin-valve effect
  from chirality-induced spin selectivity}.
\newblock \emph{\bibinfo{journal}{ACS Nano}} \textbf{\bibinfo{volume}{14}},
  \bibinfo{pages}{15983--15991} (\bibinfo{year}{2020}).
\newblock \urlprefix\url{https://pubs.acs.org/doi/10.1021/acsnano.0c07438}.

\bibitem{Kim2021}
\bibinfo{author}{Kim, Y.-H.} \emph{et~al.}
\newblock \bibinfo{title}{{Chiral-induced spin selectivity enables a
  room-temperature spin light-emitting diode}}.
\newblock \emph{\bibinfo{journal}{Science}} \textbf{\bibinfo{volume}{371}},
  \bibinfo{pages}{1129--1133} (\bibinfo{year}{2021}).
\newblock
  \urlprefix\url{https://www.science.org/doi/full/10.1126/science.abf5291}.

\bibitem{Qian2022}
\bibinfo{author}{Qian, Q.} \emph{et~al.}
\newblock \bibinfo{title}{{Chiral molecular intercalation superlattices}}.
\newblock \emph{\bibinfo{journal}{Nature}} \textbf{\bibinfo{volume}{606}},
  \bibinfo{pages}{902--908} (\bibinfo{year}{2022}).
\newblock \urlprefix\url{https://www.nature.com/articles/s41586-022-04846-3}.

\bibitem{Liu2022experiment}
\bibinfo{author}{Adhikari, Y.} \emph{et~al.}
\newblock \bibinfo{title}{Interplay of structural chirality, electron spin and
  topological orbital in chiral molecular spin valves}.
\newblock \emph{\bibinfo{journal}{Nature Communications}}
  \textbf{\bibinfo{volume}{14}}, \bibinfo{pages}{5163} (\bibinfo{year}{2023}).
\newblock \urlprefix\url{https://www.nature.com/articles/s41467-023-40884-9}.

\bibitem{Al-Bustami2022}
\bibinfo{author}{Al-Bustami, H.} \emph{et~al.}
\newblock \bibinfo{title}{Atomic and molecular layer deposition of chiral thin
  films showing up to 99\% spin selective transport}.
\newblock \emph{\bibinfo{journal}{Nano Letters}} \textbf{\bibinfo{volume}{22}},
  \bibinfo{pages}{5022--5028} (\bibinfo{year}{2022}).
\newblock \urlprefix\url{https://doi.org/10.1021/acs.nanolett.2c01953}.
\newblock \eprint{https://doi.org/10.1021/acs.nanolett.2c01953}.

\bibitem{Kulkarni2020}
\bibinfo{author}{Kulkarni, C.} \emph{et~al.}
\newblock \bibinfo{title}{{Highly Efficient and Tunable Filtering of Electrons'
  Spin by Supramolecular Chirality of Nanofiber‐Based Materials}}.
\newblock \emph{\bibinfo{journal}{Advanced Materials}}
  \textbf{\bibinfo{volume}{32}}, \bibinfo{pages}{1904965}
  (\bibinfo{year}{2020}).
\newblock
  \urlprefix\url{https://onlinelibrary.wiley.com/doi/full/10.1002/adma.201904965}.

\bibitem{Mishra2020}
\bibinfo{author}{Mishra, S.} \emph{et~al.}
\newblock \bibinfo{title}{{Length-Dependent Electron Spin Polarization in
  Oligopeptides and DNA}}.
\newblock \emph{\bibinfo{journal}{The Journal of Physical Chemistry C}}
  \textbf{\bibinfo{volume}{124}}, \bibinfo{pages}{10776--10782}
  (\bibinfo{year}{2020}).
\newblock \urlprefix\url{https://pubs.acs.org/doi/10.1021/acs.jpcc.0c02291}.

\bibitem{Yang2019}
\bibinfo{author}{Yang, X.}, \bibinfo{author}{Wal, C. H. v.~d.} \&
  \bibinfo{author}{Wees, B. J.~v.}
\newblock \bibinfo{title}{{Spin-dependent electron transmission model for
  chiral molecules in mesoscopic devices}}.
\newblock \emph{\bibinfo{journal}{Physical Review B}}
  \textbf{\bibinfo{volume}{99}}, \bibinfo{pages}{024418}
  (\bibinfo{year}{2019}).
\newblock
  \urlprefix\url{https://journals.aps.org/prb/abstract/10.1103/PhysRevB.99.024418}.
\newblock \eprint{1810.02662}.

\bibitem{Yang2019b}
\bibinfo{author}{Yang, X.}, \bibinfo{author}{van~der Wal, C.~H.} \&
  \bibinfo{author}{van Wees, B.~J.}
\newblock \bibinfo{title}{Detecting chirality in two-terminal electronic
  nanodevices}.
\newblock \emph{\bibinfo{journal}{Nano Letters}} \textbf{\bibinfo{volume}{20}},
  \bibinfo{pages}{6148--6154} (\bibinfo{year}{2020}).
\newblock
  \urlprefix\url{https://pubs.acs.org/doi/10.1021/acs.nanolett.0c02417}.

\bibitem{Naaman2020comment}
\bibinfo{author}{Naaman, R.} \& \bibinfo{author}{Waldeck, D.~H.}
\newblock \bibinfo{title}{{Comment on ``Spin-dependent electron transmission
  model for chiral molecules in mesoscopic devices''}}.
\newblock \emph{\bibinfo{journal}{Physical Review B}}
  \textbf{\bibinfo{volume}{101}}, \bibinfo{pages}{026403}
  (\bibinfo{year}{2020}).
\newblock
  \urlprefix\url{https://journals.aps.org/prb/abstract/10.1103/PhysRevB.99.024418}.

\bibitem{Yang2020reply}
\bibinfo{author}{Yang, X.}, \bibinfo{author}{van~der Wal, C.~H.} \&
  \bibinfo{author}{van Wees, B.~J.}
\newblock \bibinfo{title}{Reply to ``comment on `spin-dependent electron
  transmission model for chiral molecules in mesoscopic devices'''}.
\newblock \emph{\bibinfo{journal}{Phys. Rev. B}}
  \textbf{\bibinfo{volume}{101}}, \bibinfo{pages}{026404}
  (\bibinfo{year}{2020}).
\newblock \urlprefix\url{https://link.aps.org/doi/10.1103/PhysRevB.101.026404}.

\bibitem{Dalum2019}
\bibinfo{author}{Dalum, S.} \& \bibinfo{author}{Hedegard, P.}
\newblock \bibinfo{title}{{Theory of Chiral Induced Spin Selectivity}}.
\newblock \emph{\bibinfo{journal}{Nano Letters}} \textbf{\bibinfo{volume}{19}},
  \bibinfo{pages}{5253--5259} (\bibinfo{year}{2019}).
\newblock
  \urlprefix\url{https://pubs.acs.org/doi/10.1021/acs.nanolett.9b01707}.

\bibitem{Rikken2001}
\bibinfo{author}{Rikken, G. L. J.~A.}, \bibinfo{author}{F\"{a}lling, J.} \&
  \bibinfo{author}{Wyder, P.}
\newblock \bibinfo{title}{{Electrical Magnetochiral Anisotropy}}.
\newblock \emph{\bibinfo{journal}{Physical Review Letters}}
  \textbf{\bibinfo{volume}{87}}, \bibinfo{pages}{236602}
  (\bibinfo{year}{2001}).
\newblock
  \urlprefix\url{https://journals.aps.org/prl/pdf/10.1103/PhysRevLett.87.236602}.

\bibitem{Rikken2019}
\bibinfo{author}{Rikken, G. L. J.~A.} \& \bibinfo{author}{Avarvari, N.}
\newblock \bibinfo{title}{Strong electrical magnetochiral anisotropy in
  tellurium}.
\newblock \emph{\bibinfo{journal}{Phys. Rev. B}} \textbf{\bibinfo{volume}{99}},
  \bibinfo{pages}{245153} (\bibinfo{year}{2019}).
\newblock \urlprefix\url{https://link.aps.org/doi/10.1103/PhysRevB.99.245153}.

\bibitem{pop2014electrical}
\bibinfo{author}{Pop, F.}, \bibinfo{author}{Auban-Senzier, P.},
  \bibinfo{author}{Canadell, E.}, \bibinfo{author}{Rikken, G.~L.} \&
  \bibinfo{author}{Avarvari, N.}
\newblock \bibinfo{title}{Electrical magnetochiral anisotropy in a bulk chiral
  molecular conductor}.
\newblock \emph{\bibinfo{journal}{Nature Communications}}
  \textbf{\bibinfo{volume}{5}}, \bibinfo{pages}{3757} (\bibinfo{year}{2014}).
\newblock \urlprefix\url{https://www.nature.com/articles/ncomms4757}.

\bibitem{Aoki2019}
\bibinfo{author}{Aoki, R.}, \bibinfo{author}{Kousaka, Y.} \&
  \bibinfo{author}{Togawa, Y.}
\newblock \bibinfo{title}{Anomalous nonreciprocal electrical transport on
  chiral magnetic order}.
\newblock \emph{\bibinfo{journal}{Phys. Rev. Lett.}}
  \textbf{\bibinfo{volume}{122}}, \bibinfo{pages}{057206}
  (\bibinfo{year}{2019}).
\newblock
  \urlprefix\url{https://link.aps.org/doi/10.1103/PhysRevLett.122.057206}.

\bibitem{Inui2020}
\bibinfo{author}{Inui, A.} \emph{et~al.}
\newblock \bibinfo{title}{Chirality-induced spin-polarized state of a chiral
  crystal ${\mathrm{crnb}}_{3}{\mathrm{s}}_{6}$}.
\newblock \emph{\bibinfo{journal}{Phys. Rev. Lett.}}
  \textbf{\bibinfo{volume}{124}}, \bibinfo{pages}{166602}
  (\bibinfo{year}{2020}).
\newblock
  \urlprefix\url{https://link.aps.org/doi/10.1103/PhysRevLett.124.166602}.

\bibitem{Shiota2021}
\bibinfo{author}{Shiota, K.} \emph{et~al.}
\newblock \bibinfo{title}{Chirality-induced spin polarization over macroscopic
  distances in chiral disilicide crystals}.
\newblock \emph{\bibinfo{journal}{Phys. Rev. Lett.}}
  \textbf{\bibinfo{volume}{127}}, \bibinfo{pages}{126602}
  (\bibinfo{year}{2021}).
\newblock
  \urlprefix\url{https://link.aps.org/doi/10.1103/PhysRevLett.127.126602}.

\bibitem{Ye2022Te}
\bibinfo{author}{Niu, C.} \emph{et~al.}
\newblock \bibinfo{title}{{Tunable nonreciprocal electrical transport in 2D
  Tellurium with different chirality}}.
\newblock \emph{\bibinfo{journal}{arXiv}}  (\bibinfo{year}{2022}).
\newblock \eprint{2201.08829}.

\bibitem{Onsager1931b}
\bibinfo{author}{Onsager, L.}
\newblock \bibinfo{title}{Reciprocal relations in irreversible processes. ii.}
\newblock \emph{\bibinfo{journal}{Phys. Rev.}} \textbf{\bibinfo{volume}{38}},
  \bibinfo{pages}{2265--2279} (\bibinfo{year}{1931}).
\newblock \urlprefix\url{https://link.aps.org/doi/10.1103/PhysRev.38.2265}.

\bibitem{LANDAU1980}
\bibinfo{author}{Landau, L.} \& \bibinfo{author}{Lifshitz, E.}
\newblock \bibinfo{title}{Chapter xii - fluctuations}.
\newblock In \bibinfo{editor}{Landau, L.} \& \bibinfo{editor}{Lifshitz, E.}
  (eds.) \emph{\bibinfo{booktitle}{Statistical Physics (Third Edition)}},
  \bibinfo{pages}{333--400} (\bibinfo{publisher}{Butterworth-Heinemann},
  \bibinfo{address}{Oxford}, \bibinfo{year}{1980}), \bibinfo{edition}{third
  edition} edn.
\newblock
  \urlprefix\url{https://www.sciencedirect.com/science/article/pii/B9780080570464500191}.

\bibitem{Buttiker1988}
\bibinfo{author}{B\"uttiker, M.}
\newblock \bibinfo{title}{{Symmetry of electrical conduction}}.
\newblock \emph{\bibinfo{journal}{IBM Journal of Research and Development}}
  \textbf{\bibinfo{volume}{32}}, \bibinfo{pages}{317--334}
  (\bibinfo{year}{1988}).
\newblock \urlprefix\url{https://ieeexplore.ieee.org/document/5390001}.

\bibitem{Ideue2017}
\bibinfo{author}{Ideue, T.} \emph{et~al.}
\newblock \bibinfo{title}{{Bulk rectification effect in a polar
  semiconductor}}.
\newblock \emph{\bibinfo{journal}{Nature Physics}}
  \textbf{\bibinfo{volume}{13}}, \bibinfo{pages}{578--583}
  (\bibinfo{year}{2017}).

\bibitem{Liu2021UMR}
\bibinfo{author}{Liu, Y.}, \bibinfo{author}{Holder, T.} \&
  \bibinfo{author}{Yan, B.}
\newblock \bibinfo{title}{Chirality-induced giant unidirectional
  magnetoresistance in twisted bilayer graphene}.
\newblock \emph{\bibinfo{journal}{The Innovation}}
  \textbf{\bibinfo{volume}{2}}, \bibinfo{pages}{100085} (\bibinfo{year}{2021}).
\newblock
  \urlprefix\url{https://www.sciencedirect.com/science/article/pii/S2666675821000102}.

\bibitem{kaplan2022unification}
\bibinfo{author}{Kaplan, D.}, \bibinfo{author}{Holder, T.} \&
  \bibinfo{author}{Yan, B.}
\newblock \bibinfo{title}{Unification of nonlinear anomalous hall effect and
  nonreciprocal magnetoresistance in metals by the quantum geometry}.
\newblock \emph{\bibinfo{journal}{Phys. Rev. Lett.}}
  \textbf{\bibinfo{volume}{132}}, \bibinfo{pages}{026301}
  (\bibinfo{year}{2024}).
\newblock
  \urlprefix\url{https://link.aps.org/doi/10.1103/PhysRevLett.132.026301}.

\bibitem{Wolf2022}
\bibinfo{author}{Wolf, Y.}, \bibinfo{author}{Liu, Y.}, \bibinfo{author}{Xiao,
  J.}, \bibinfo{author}{Park, N.} \& \bibinfo{author}{Yan, B.}
\newblock \bibinfo{title}{{Unusual Spin Polarization in the Chirality-Induced
  Spin Selectivity}}.
\newblock \emph{\bibinfo{journal}{ACS Nano}} \textbf{\bibinfo{volume}{16}},
  \bibinfo{pages}{18601--18607} (\bibinfo{year}{2022}).
\newblock \urlprefix\url{https://pubs.acs.org/doi/10.1021/acsnano.2c07088}.

\bibitem{Naskar2023}
\bibinfo{author}{Naskar, S.}, \bibinfo{author}{Mujica, V.} \&
  \bibinfo{author}{Herrmann, C.}
\newblock \bibinfo{title}{{Chiral-Induced Spin Selectivity and Non-equilibrium
  Spin Accumulation in Molecules and Interfaces: A First-Principles Study}}.
\newblock \emph{\bibinfo{journal}{The Journal of Physical Chemistry Letters}}
  \textbf{\bibinfo{volume}{14}}, \bibinfo{pages}{694--701}
  (\bibinfo{year}{2023}).
\newblock \urlprefix\url{https://pubs.acs.org/doi/10.1021/acs.jpclett.2c03747}.

\bibitem{adhikari2023interplay}
\bibinfo{author}{Adhikari, Y.} \emph{et~al.}
\newblock \bibinfo{title}{Interplay of structural chirality, electron spin and
  topological orbital in chiral molecular spin valves}.
\newblock \emph{\bibinfo{journal}{Nature Communications}}
  \textbf{\bibinfo{volume}{14}}, \bibinfo{pages}{5163} (\bibinfo{year}{2023}).
\newblock \urlprefix\url{https://www.nature.com/articles/s41467-023-40884-9}.

\bibitem{Dubi2021}
\bibinfo{author}{Alwan, S.} \& \bibinfo{author}{Dubi, Y.}
\newblock \bibinfo{title}{{Spinterface Origin for the Chirality-Induced
  Spin-Selectivity Effect}}.
\newblock \emph{\bibinfo{journal}{Journal of the American Chemical Society}}
  \textbf{\bibinfo{volume}{143}}, \bibinfo{pages}{14235--14241}
  (\bibinfo{year}{2021}).
\newblock \urlprefix\url{https://pubs.acs.org/doi/10.1021/jacs.1c05637}.

\bibitem{Coropceanu2007}
\bibinfo{author}{Coropceanu, V.} \emph{et~al.}
\newblock \bibinfo{title}{{Charge Transport in Organic Semiconductors}}.
\newblock \emph{\bibinfo{journal}{Chemical Reviews}}
  \textbf{\bibinfo{volume}{107}}, \bibinfo{pages}{926--952}
  (\bibinfo{year}{2007}).

\bibitem{Kaake2010}
\bibinfo{author}{Kaake, L.~G.}, \bibinfo{author}{Barbara, P.~F.} \&
  \bibinfo{author}{Zhu, X.-Y.}
\newblock \bibinfo{title}{{Intrinsic Charge Trapping in Organic and Polymeric
  Semiconductors: A Physical Chemistry Perspective}}.
\newblock \emph{\bibinfo{journal}{The Journal of Physical Chemistry Letters}}
  \textbf{\bibinfo{volume}{1}}, \bibinfo{pages}{628--635}
  (\bibinfo{year}{2010}).

\bibitem{Haneef2019}
\bibinfo{author}{Haneef, H.~F.}, \bibinfo{author}{Zeidell, A.~M.} \&
  \bibinfo{author}{Jurchescu, O.~D.}
\newblock \bibinfo{title}{{Charge carrier traps in organic semiconductors: a
  review on the underlying physics and impact on electronic devices}}.
\newblock \emph{\bibinfo{journal}{Journal of Materials Chemistry C}}
  \textbf{\bibinfo{volume}{8}}, \bibinfo{pages}{759--787}
  (\bibinfo{year}{2019}).

\bibitem{sachnik2023elimination}
\bibinfo{author}{Sachnik, O.} \emph{et~al.}
\newblock \bibinfo{title}{Elimination of charge-carrier trapping by molecular
  design}.
\newblock \emph{\bibinfo{journal}{Nature Materials}}
  \textbf{\bibinfo{volume}{22}}, \bibinfo{pages}{1114--1120}
  (\bibinfo{year}{2023}).

\bibitem{Novembre2008memory}
\bibinfo{author}{Novembre, C.}, \bibinfo{author}{Guérin, D.},
  \bibinfo{author}{Lmimouni, K.}, \bibinfo{author}{Gamrat, C.} \&
  \bibinfo{author}{Vuillaume, D.}
\newblock \bibinfo{title}{{Gold nanoparticle-pentacene memory transistors}}.
\newblock \emph{\bibinfo{journal}{Applied Physics Letters}}
  \textbf{\bibinfo{volume}{92}}, \bibinfo{pages}{103314}
  (\bibinfo{year}{2008}).
\newblock \eprint{0802.2633}.

\bibitem{Son2009memory}
\bibinfo{author}{Son, D.~I.}, \bibinfo{author}{You, C.~H.},
  \bibinfo{author}{Kim, W.~T.}, \bibinfo{author}{Jung, J.~H.} \&
  \bibinfo{author}{Kim, T.~W.}
\newblock \bibinfo{title}{{Electrical bistabilities and memory mechanisms of
  organic bistable devices based on colloidal ZnO quantum
  dot-polymethylmethacrylate polymer nanocomposites}}.
\newblock \emph{\bibinfo{journal}{Applied Physics Letters}}
  \textbf{\bibinfo{volume}{94}}, \bibinfo{pages}{132103}
  (\bibinfo{year}{2009}).

\bibitem{Burgt2018}
\bibinfo{author}{Burgt, Y. v.~d.}, \bibinfo{author}{Melianas, A.},
  \bibinfo{author}{Keene, S.~T.}, \bibinfo{author}{Malliaras, G.} \&
  \bibinfo{author}{Salleo, A.}
\newblock \bibinfo{title}{{Organic electronics for neuromorphic computing}}.
\newblock \emph{\bibinfo{journal}{Nature Electronics}}
  \textbf{\bibinfo{volume}{1}}, \bibinfo{pages}{386--397}
  (\bibinfo{year}{2018}).

\bibitem{Park2022}
\bibinfo{author}{Park, J.} \emph{et~al.}
\newblock \bibinfo{title}{Controlled hysteresis of conductance in molecular
  tunneling junctions}.
\newblock \emph{\bibinfo{journal}{ACS Nano}} \textbf{\bibinfo{volume}{16}},
  \bibinfo{pages}{4206--4216} (\bibinfo{year}{2022}).
\newblock \urlprefix\url{https://doi.org/10.1021/acsnano.1c10155}.

\bibitem{Yao2018}
\bibinfo{author}{Yao, S.} \& \bibinfo{author}{Wang, Z.}
\newblock \bibinfo{title}{{Edge States and Topological Invariants of
  Non-Hermitian Systems}}.
\newblock \emph{\bibinfo{journal}{Phys. Rev. Lett.}}
  \textbf{\bibinfo{volume}{121}}, \bibinfo{pages}{086803}
  (\bibinfo{year}{2018}).
\newblock
  \urlprefix\url{https://journals.aps.org/prl/abstract/10.1103/PhysRevLett.121.086803}.

\bibitem{Murakami2019PRL}
\bibinfo{author}{Yokomizo, K.} \& \bibinfo{author}{Murakami, S.}
\newblock \bibinfo{title}{{Non-Bloch Band Theory of Non-Hermitian Systems}}.
\newblock \emph{\bibinfo{journal}{Phys. Rev. Lett.}}
  \textbf{\bibinfo{volume}{123}}, \bibinfo{pages}{066404}
  (\bibinfo{year}{2019}).
\newblock
  \urlprefix\url{https://journals.aps.org/prl/abstract/10.1103/PhysRevLett.123.066404}.

\bibitem{Zhang2020}
\bibinfo{author}{Zhang, K.}, \bibinfo{author}{Yang, Z.} \&
  \bibinfo{author}{Fang, C.}
\newblock \bibinfo{title}{Correspondence between winding numbers and skin modes
  in non-hermitian systems}.
\newblock \emph{\bibinfo{journal}{Phys. Rev. Lett.}}
  \textbf{\bibinfo{volume}{125}}, \bibinfo{pages}{126402}
  (\bibinfo{year}{2020}).
\newblock
  \urlprefix\url{https://link.aps.org/doi/10.1103/PhysRevLett.125.126402}.

\bibitem{YYF2020PRL}
\bibinfo{author}{Yi, Y.} \& \bibinfo{author}{Yang, Z.}
\newblock \bibinfo{title}{Non-hermitian skin modes induced by on-site
  dissipations and chiral tunneling effect}.
\newblock \emph{\bibinfo{journal}{Phys. Rev. Lett.}}
  \textbf{\bibinfo{volume}{125}}, \bibinfo{pages}{186802}
  (\bibinfo{year}{2020}).
\newblock
  \urlprefix\url{https://link.aps.org/doi/10.1103/PhysRevLett.125.186802}.

\bibitem{OkumaSatoReview}
\bibinfo{author}{Okuma, N.} \& \bibinfo{author}{Sato, M.}
\newblock \bibinfo{title}{Non-hermitian topological phenomena: A review}.
\newblock \emph{\bibinfo{journal}{Annual Review of Condensed Matter Physics}}
  \textbf{\bibinfo{volume}{14}}, \bibinfo{pages}{83--107}
  (\bibinfo{year}{2023}).
\newblock
  \urlprefix\url{https://www.annualreviews.org/content/journals/10.1146/annurev-conmatphys-040521-033133}.

\bibitem{LeeCHReview}
\bibinfo{author}{Lin, R.}, \bibinfo{author}{Tai, T.}, \bibinfo{author}{Li, L.}
  \& \bibinfo{author}{Lee, C.~H.}
\newblock \bibinfo{title}{Topological non-hermitian skin effect}.
\newblock \emph{\bibinfo{journal}{Frontiers of Physics}}
  \textbf{\bibinfo{volume}{18}}, \bibinfo{pages}{53605} (\bibinfo{year}{2023}).
\newblock \urlprefix\url{https://doi.org/10.1007/s11467-023-1309-z}.

\bibitem{Sato2019PRX}
\bibinfo{author}{Kawabata, K.}, \bibinfo{author}{Shiozaki, K.},
  \bibinfo{author}{Ueda, M.} \& \bibinfo{author}{Sato, M.}
\newblock \bibinfo{title}{Symmetry and topology in non-hermitian physics}.
\newblock \emph{\bibinfo{journal}{Phys. Rev. X}} \textbf{\bibinfo{volume}{9}},
  \bibinfo{pages}{041015} (\bibinfo{year}{2019}).
\newblock \urlprefix\url{https://link.aps.org/doi/10.1103/PhysRevX.9.041015}.

\bibitem{XuePeng2020}
\bibinfo{author}{Xiao, L.} \emph{et~al.}
\newblock \bibinfo{title}{{Non-Hermitian bulk--boundary correspondence in
  quantum dynamics}}.
\newblock \emph{\bibinfo{journal}{Nature Physics}}
  \textbf{\bibinfo{volume}{16}}, \bibinfo{pages}{761--766}
  (\bibinfo{year}{2020}).

\bibitem{Thomale2020}
\bibinfo{author}{Helbig, T.} \emph{et~al.}
\newblock \bibinfo{title}{{Generalized bulk--boundary correspondence in
  non-Hermitian topolectrical circuits}}.
\newblock \emph{\bibinfo{journal}{Nature Physics}}
  \textbf{\bibinfo{volume}{16}}, \bibinfo{pages}{747--750}
  (\bibinfo{year}{2020}).

\bibitem{Ghatak2020}
\bibinfo{author}{Ghatak, A.}, \bibinfo{author}{Brandenbourger, M.},
  \bibinfo{author}{van Wezel, J.} \& \bibinfo{author}{Coulais, C.}
\newblock \bibinfo{title}{{Observation of non-Hermitian topology and its
  bulk{\textendash}edge correspondence in an active mechanical metamaterial}}.
\newblock \emph{\bibinfo{journal}{Proceedings of the National Academy of
  Sciences}} \textbf{\bibinfo{volume}{117}}, \bibinfo{pages}{29561--29568}
  (\bibinfo{year}{2020}).

\bibitem{Fulga2024NP}
\bibinfo{author}{Ochkan, K.} \emph{et~al.}
\newblock \bibinfo{title}{Non-hermitian topology in a multi-terminal quantum
  hall device}.
\newblock \emph{\bibinfo{journal}{Nature Physics}}  (\bibinfo{year}{2024}).
\newblock \urlprefix\url{https://doi.org/10.1038/s41567-023-02337-4}.

\bibitem{Abendroth2019}
\bibinfo{author}{Abendroth, J.~M.} \emph{et~al.}
\newblock \bibinfo{title}{{Spin-Dependent Ionization of Chiral Molecular
  Films}}.
\newblock \emph{\bibinfo{journal}{Journal of the American Chemical Society}}
  \textbf{\bibinfo{volume}{141}}, \bibinfo{pages}{3863--3874}
  (\bibinfo{year}{2019}).
\newblock \urlprefix\url{https://pubs.acs.org/doi/10.1021/jacs.8b08421}.

\bibitem{Ghosh2020}
\bibinfo{author}{Ghosh, S.} \emph{et~al.}
\newblock \bibinfo{title}{{Effect of Chiral Molecules on the Electrons' Spin
  Wavefunction at Interfaces}}.
\newblock \emph{\bibinfo{journal}{The Journal of Physical Chemistry Letters}}
  \textbf{\bibinfo{volume}{11}}, \bibinfo{pages}{1550--1557}
  (\bibinfo{year}{2020}).
\newblock \urlprefix\url{https://pubs.acs.org/doi/10.1021/acs.jpclett.9b03487}.

\bibitem{Wu2024gold}
\bibinfo{author}{Wu, F.} \emph{et~al.}
\newblock \bibinfo{title}{{Enantiomer-Selective Magnetoresistance in Chiral
  Gold Nanocrystals by Magnetic Control of Surface Potentials}}.
\newblock \emph{\bibinfo{journal}{arXiv:2408.03501}}  (\bibinfo{year}{2024}).
\newblock \urlprefix\url{https://arxiv.org/abs/2408.03501}.

\bibitem{Verhage2024crystal}
\bibinfo{author}{Verhage, M.} \emph{et~al.}
\newblock \bibinfo{title}{{Chirality‐Induced Magnetic Polarization by Charge
  Localization in a Chiral Supramolecular Crystal}}.
\newblock \emph{\bibinfo{journal}{Advanced Materials}}
  \textbf{\bibinfo{volume}{36}}, \bibinfo{pages}{e2403807}
  (\bibinfo{year}{2024}).
\newblock
  \urlprefix\url{https://onlinelibrary.wiley.com/doi/10.1002/adma.202403807}.

\bibitem{Banerjee2020}
\bibinfo{author}{Banerjee-Ghosh, K.} \emph{et~al.}
\newblock \bibinfo{title}{{Long-Range Charge Reorganization as an Allosteric
  Control Signal in Proteins}}.
\newblock \emph{\bibinfo{journal}{Journal of the American Chemical Society}}
  \textbf{\bibinfo{volume}{142}}, \bibinfo{pages}{20456--20462}
  (\bibinfo{year}{2020}).
\newblock \urlprefix\url{https://pubs.acs.org/doi/10.1021/jacs.0c10105}.

\bibitem{Ghosh2021}
\bibinfo{author}{Ghosh, S.} \emph{et~al.}
\newblock \bibinfo{title}{{Substrates Modulate Charge-Reorganization Allosteric
  Effects in Protein–Protein Association}}.
\newblock \emph{\bibinfo{journal}{The Journal of Physical Chemistry Letters}}
  \textbf{\bibinfo{volume}{12}}, \bibinfo{pages}{2805--2808}
  (\bibinfo{year}{2021}).

\bibitem{Naaman2022}
\bibinfo{author}{Naaman, R.}, \bibinfo{author}{Paltiel, Y.} \&
  \bibinfo{author}{Waldeck, D.~H.}
\newblock \bibinfo{title}{{Chiral Induced Spin Selectivity and Its Implications
  for Biological Functions}}.
\newblock \emph{\bibinfo{journal}{Annual Review of Biophysics}}
  \textbf{\bibinfo{volume}{51}}, \bibinfo{pages}{99--114}
  (\bibinfo{year}{2022}).
\newblock
  \urlprefix\url{https://www.annualreviews.org/doi/10.1146/annurev-biophys-083021-070400}.

\bibitem{lu2023beyond}
\bibinfo{author}{Lu, Y.}, \bibinfo{author}{Joy, M.}, \bibinfo{author}{Bloom,
  B.~P.} \& \bibinfo{author}{Waldeck, D.~H.}
\newblock \bibinfo{title}{Beyond stereoisomeric effects: Exploring the
  importance of intermolecular electron spin interactions in biorecognition}.
\newblock \emph{\bibinfo{journal}{The Journal of Physical Chemistry Letters}}
  \textbf{\bibinfo{volume}{14}}, \bibinfo{pages}{7032--7037}
  (\bibinfo{year}{2023}).

\bibitem{zuo2023mechano}
\bibinfo{author}{Zuo, L.} \emph{et~al.}
\newblock \bibinfo{title}{Mechano-electron spin coupling modulates the
  reactivity of individual coronazymes}.
\newblock \emph{\bibinfo{journal}{ChemRxiv}}  (\bibinfo{year}{2023}).

\bibitem{Liu2023spin}
\bibinfo{author}{Liu, T.} \& \bibinfo{author}{Weiss, P.~S.}
\newblock \bibinfo{title}{Spin polarization in transport studies of
  chirality-induced spin selectivity}.
\newblock \emph{\bibinfo{journal}{ACS Nano}} \textbf{\bibinfo{volume}{17}},
  \bibinfo{pages}{19502--19507} (\bibinfo{year}{2023}).
\newblock \urlprefix\url{https://doi.org/10.1021/acsnano.3c06133}.

\bibitem{Kawabata2020PRB}
\bibinfo{author}{Kawabata, K.}, \bibinfo{author}{Okuma, N.} \&
  \bibinfo{author}{Sato, M.}
\newblock \bibinfo{title}{Non-bloch band theory of non-hermitian hamiltonians
  in the symplectic class}.
\newblock \emph{\bibinfo{journal}{Phys. Rev. B}}
  \textbf{\bibinfo{volume}{101}}, \bibinfo{pages}{195147}
  (\bibinfo{year}{2020}).
\newblock \urlprefix\url{https://link.aps.org/doi/10.1103/PhysRevB.101.195147}.

\bibitem{Guo2012}
\bibinfo{author}{Guo, A.-M.} \& \bibinfo{author}{Sun, Q.-f.}
\newblock \bibinfo{title}{{Spin-Selective Transport of Electrons in DNA Double
  Helix}}.
\newblock \emph{\bibinfo{journal}{Physical Review Letters}}
  \textbf{\bibinfo{volume}{108}}, \bibinfo{pages}{218102}
  (\bibinfo{year}{2012}).
\newblock \eprint{1201.4888}.

\bibitem{Guo2014a}
\bibinfo{author}{Guo, A.-M.} \emph{et~al.}
\newblock \bibinfo{title}{Contact effects in spin transport along
  double-helical molecules}.
\newblock \emph{\bibinfo{journal}{Phys. Rev. B}} \textbf{\bibinfo{volume}{89}},
  \bibinfo{pages}{205434} (\bibinfo{year}{2014}).
\newblock \urlprefix\url{https://link.aps.org/doi/10.1103/PhysRevB.89.205434}.

\bibitem{Matityahu2016}
\bibinfo{author}{Matityahu, S.}, \bibinfo{author}{Utsumi, Y.},
  \bibinfo{author}{Aharony, A.}, \bibinfo{author}{Entin-Wohlman, O.} \&
  \bibinfo{author}{Balseiro, C.~A.}
\newblock \bibinfo{title}{{Spin-dependent transport through a chiral molecule
  in the presence of spin-orbit interaction and nonunitary effects}}.
\newblock \emph{\bibinfo{journal}{Physical Review B}}
  \textbf{\bibinfo{volume}{93}}, \bibinfo{pages}{075407}
  (\bibinfo{year}{2016}).

\bibitem{Piccinin2003}
\bibinfo{author}{Piccinin, S.}, \bibinfo{author}{Selloni, A.},
  \bibinfo{author}{Scandolo, S.}, \bibinfo{author}{Car, R.} \&
  \bibinfo{author}{Scoles, G.}
\newblock \bibinfo{title}{{Electronic properties of metal–molecule–metal
  systems at zero bias: A periodic density functional study}}.
\newblock \emph{\bibinfo{journal}{The Journal of Chemical Physics}}
  \textbf{\bibinfo{volume}{119}}, \bibinfo{pages}{6729--6735}
  (\bibinfo{year}{2003}).

\bibitem{Souza2014}
\bibinfo{author}{Souza, A. d.~M.} \emph{et~al.}
\newblock \bibinfo{title}{{Stretching of BDT-gold molecular junctions: thiol or
  thiolate termination?}}
\newblock \emph{\bibinfo{journal}{Nanoscale}} \textbf{\bibinfo{volume}{6}},
  \bibinfo{pages}{14495--14507} (\bibinfo{year}{2014}).

\bibitem{Zhuo2009oxygen}
\bibinfo{author}{Zhuo, J.} \emph{et~al.}
\newblock \bibinfo{title}{{Direct Spectroscopic Evidence for a Photodoping
  Mechanism in Polythiophene and Poly(bithiophene‐alt‐thienothiophene)
  Organic Semiconductor Thin Films Involving Oxygen and Sorbed Moisture}}.
\newblock \emph{\bibinfo{journal}{Advanced Materials}}
  \textbf{\bibinfo{volume}{21}}, \bibinfo{pages}{4747--4752}
  (\bibinfo{year}{2009}).

\bibitem{Nikolka2019water}
\bibinfo{author}{Nikolka, M.} \emph{et~al.}
\newblock \bibinfo{title}{{High-mobility, trap-free charge transport in
  conjugated polymer diodes}}.
\newblock \emph{\bibinfo{journal}{Nature Communications}}
  \textbf{\bibinfo{volume}{10}}, \bibinfo{pages}{2122} (\bibinfo{year}{2019}).

\bibitem{Kotadiya2019water}
\bibinfo{author}{Kotadiya, N.~B.}, \bibinfo{author}{Mondal, A.},
  \bibinfo{author}{Blom, P. W.~M.}, \bibinfo{author}{Andrienko, D.} \&
  \bibinfo{author}{Wetzelaer, G.-J. A.~H.}
\newblock \bibinfo{title}{{A window to trap-free charge transport in organic
  semiconducting thin films}}.
\newblock \emph{\bibinfo{journal}{Nature Materials}}
  \textbf{\bibinfo{volume}{18}}, \bibinfo{pages}{1182--1186}
  (\bibinfo{year}{2019}).

\bibitem{Akkerman2007}
\bibinfo{author}{Akkerman, H.~B.} \emph{et~al.}
\newblock \bibinfo{title}{{Electron tunneling through alkanedithiol
  self-assembled monolayers in large-area molecular junctions}}.
\newblock \emph{\bibinfo{journal}{Proceedings of the National Academy of
  Sciences}} \textbf{\bibinfo{volume}{104}}, \bibinfo{pages}{11161--11166}
  (\bibinfo{year}{2007}).
\newblock \urlprefix\url{https://www.pnas.org/doi/10.1073/pnas.0701472104}.

\bibitem{rikken2023comparing}
\bibinfo{author}{Rikken, G.} \& \bibinfo{author}{Avarvari, N.}
\newblock \bibinfo{title}{Comparing electrical magnetochiral anisotropy and
  chirality-induced spin selectivity}.
\newblock \emph{\bibinfo{journal}{The Journal of Physical Chemistry Letters}}
  \textbf{\bibinfo{volume}{14}}, \bibinfo{pages}{9727--9731}
  (\bibinfo{year}{2023}).
\newblock
  \urlprefix\url{https://pubs.acs.org/doi/full/10.1021/acs.jpclett.3c02546}.

\bibitem{Simmons1963}
\bibinfo{author}{Simmons, J.~G.}
\newblock \bibinfo{title}{{Generalized Formula for the Electric Tunnel Effect
  between Similar Electrodes Separated by a Thin Insulating Film}}.
\newblock \emph{\bibinfo{journal}{Journal of Applied Physics}}
  \textbf{\bibinfo{volume}{34}}, \bibinfo{pages}{1793--1803}
  (\bibinfo{year}{1963}).
\newblock
  \urlprefix\url{https://pubs.aip.org/aip/jap/article/34/6/1793/362794/Generalized-Formula-for-the-Electric-Tunnel-Effect}.

\bibitem{Naaman2015}
\bibinfo{author}{Naaman, R.} \& \bibinfo{author}{Waldeck, D.~H.}
\newblock \bibinfo{title}{{Spintronics and Chirality: Spin Selectivity in
  Electron Transport Through Chiral Molecules}}.
\newblock \emph{\bibinfo{journal}{Annual Review of Physical Chemistry}}
  \textbf{\bibinfo{volume}{66}}, \bibinfo{pages}{263 -- 281}
  (\bibinfo{year}{2015}).
\newblock
  \urlprefix\url{https://www.annualreviews.org/doi/abs/10.1146/annurev-physchem-040214-121554}.

\bibitem{Julliere1975TMR}
\bibinfo{author}{Julliere, M.}
\newblock \bibinfo{title}{{Tunneling between ferromagnetic films}}.
\newblock \emph{\bibinfo{journal}{Physics Letters A}}
  \textbf{\bibinfo{volume}{54}}, \bibinfo{pages}{225--226}
  (\bibinfo{year}{1975}).

\bibitem{Bistritzer2011}
\bibinfo{author}{Bistritzer, R.} \& \bibinfo{author}{MacDonald, A.~H.}
\newblock \bibinfo{title}{{Moire bands in twisted double-layer graphene}}.
\newblock \emph{\bibinfo{journal}{Proceedings of the National Academy of
  Sciences}} \textbf{\bibinfo{volume}{108}} (\bibinfo{year}{2011}).
\newblock \urlprefix\url{https://www.pnas.org/content/108/30/12233}.
\newblock \eprint{1009.4203}.

\bibitem{Cao2018Mott}
\bibinfo{author}{Cao, Y.} \emph{et~al.}
\newblock \bibinfo{title}{{Correlated insulator behaviour at half-filling in
  magic-angle graphene superlattices}}.
\newblock \emph{\bibinfo{journal}{Nature}} \textbf{\bibinfo{volume}{556}},
  \bibinfo{pages}{80--84} (\bibinfo{year}{2018}).
\newblock \urlprefix\url{https://www.nature.com/articles/nature26154}.

\bibitem{Cao2018SC}
\bibinfo{author}{Cao, Y.} \emph{et~al.}
\newblock \bibinfo{title}{Unconventional superconductivity in magic-angle
  graphene superlattices}.
\newblock \emph{\bibinfo{journal}{Nature}} \textbf{\bibinfo{volume}{556}},
  \bibinfo{pages}{43--50} (\bibinfo{year}{2018}).
\newblock \urlprefix\url{https://www.nature.com/articles/nature26160}.

\bibitem{Wu2019tmd}
\bibinfo{author}{Wu, F.}, \bibinfo{author}{Lovorn, T.}, \bibinfo{author}{Tutuc,
  E.}, \bibinfo{author}{Martin, I.} \& \bibinfo{author}{MacDonald, A.~H.}
\newblock \bibinfo{title}{Topological insulators in twisted transition metal
  dichalcogenide homobilayers}.
\newblock \emph{\bibinfo{journal}{Phys. Rev. Lett.}}
  \textbf{\bibinfo{volume}{122}}, \bibinfo{pages}{086402}
  (\bibinfo{year}{2019}).
\newblock
  \urlprefix\url{https://link.aps.org/doi/10.1103/PhysRevLett.122.086402}.

\bibitem{Zhang2020tmd}
\bibinfo{author}{Zhang, Y.}, \bibinfo{author}{Yuan, N. F.~Q.} \&
  \bibinfo{author}{Fu, L.}
\newblock \bibinfo{title}{Moir\'e quantum chemistry: Charge transfer in
  transition metal dichalcogenide superlattices}.
\newblock \emph{\bibinfo{journal}{Phys. Rev. B}}
  \textbf{\bibinfo{volume}{102}}, \bibinfo{pages}{201115}
  (\bibinfo{year}{2020}).
\newblock \urlprefix\url{https://link.aps.org/doi/10.1103/PhysRevB.102.201115}.

\bibitem{Ghiotto2021tmd}
\bibinfo{author}{Ghiotto, A.} \emph{et~al.}
\newblock \bibinfo{title}{Quantum criticality in twisted transition metal
  dichalcogenides}.
\newblock \emph{\bibinfo{journal}{Nature}} \textbf{\bibinfo{volume}{597}},
  \bibinfo{pages}{345--349} (\bibinfo{year}{2021}).
\newblock \urlprefix\url{https://doi.org/10.1038/s41586-021-03815-6}.

\bibitem{Shan.Xu.2022}
\bibinfo{author}{Xu, Y.} \emph{et~al.}
\newblock \bibinfo{title}{{A tunable bilayer Hubbard model in twisted WSe2}}.
\newblock \emph{\bibinfo{journal}{Nature Nanotechnology}}
  \textbf{\bibinfo{volume}{17}}, \bibinfo{pages}{934--939}
  (\bibinfo{year}{2022}).
\newblock \urlprefix\url{https://www.nature.com/articles/s41565-022-01180-7}.

\bibitem{Wang2020}
\bibinfo{author}{Wang, C.}, \bibinfo{author}{Gao, Y.}, \bibinfo{author}{Lv,
  H.}, \bibinfo{author}{Xu, X.} \& \bibinfo{author}{Xiao, D.}
\newblock \bibinfo{title}{Stacking domain wall magnons in twisted van der waals
  magnets}.
\newblock \emph{\bibinfo{journal}{Phys. Rev. Lett.}}
  \textbf{\bibinfo{volume}{125}}, \bibinfo{pages}{247201}
  (\bibinfo{year}{2020}).
\newblock
  \urlprefix\url{https://link.aps.org/doi/10.1103/PhysRevLett.125.247201}.

\bibitem{Lian2020}
\bibinfo{author}{Lian, B.}, \bibinfo{author}{Liu, Z.}, \bibinfo{author}{Zhang,
  Y.} \& \bibinfo{author}{Wang, J.}
\newblock \bibinfo{title}{Flat chern band from twisted bilayer
  ${\mathrm{mnbi}}_{2}{\mathrm{te}}_{4}$}.
\newblock \emph{\bibinfo{journal}{Phys. Rev. Lett.}}
  \textbf{\bibinfo{volume}{124}}, \bibinfo{pages}{126402}
  (\bibinfo{year}{2020}).
\newblock
  \urlprefix\url{https://link.aps.org/doi/10.1103/PhysRevLett.124.126402}.

\bibitem{kim2020observation}
\bibinfo{author}{Kim, J.} \emph{et~al.}
\newblock \bibinfo{title}{Observation of plateau-like magnetoresistance in
  twisted fe3gete2/fe3gete2 junction}.
\newblock \emph{\bibinfo{journal}{Journal of Applied Physics}}
  \textbf{\bibinfo{volume}{128}}, \bibinfo{pages}{093901}
  (\bibinfo{year}{2020}).
\newblock \urlprefix\url{https://aip.scitation.org/doi/full/10.1063/5.0012305}.

\bibitem{Inbar2023}
\bibinfo{author}{Inbar, A.} \emph{et~al.}
\newblock \bibinfo{title}{{The quantum twisting microscope}}.
\newblock \emph{\bibinfo{journal}{Nature}} \textbf{\bibinfo{volume}{614}},
  \bibinfo{pages}{682--687} (\bibinfo{year}{2023}).
\newblock \urlprefix\url{https://www.nature.com/articles/s41586-022-05685-y}.

\bibitem{Alpern2016}
\bibinfo{author}{Alpern, H.} \emph{et~al.}
\newblock \bibinfo{title}{Unconventional superconductivity induced in nb films
  by adsorbed chiral molecules}.
\newblock \emph{\bibinfo{journal}{New Journal of Physics}}
  \textbf{\bibinfo{volume}{18}}, \bibinfo{pages}{113048}
  (\bibinfo{year}{2016}).
\newblock \urlprefix\url{https://doi.org/10.1088/1367-2630/18/11/113048}.

\bibitem{singh2024single}
\bibinfo{author}{Singh, A.~K.} \emph{et~al.}
\newblock \bibinfo{title}{Single-molecule junctions map the interplay between
  electrons and chirality}.
\newblock \emph{\bibinfo{journal}{arXiv preprint arXiv:2408.12258}}
  (\bibinfo{year}{2024}).
\newblock \urlprefix\url{https://arxiv.org/abs/2408.12258}.

\end{thebibliography}
%\bibliographystyle{naturemag}

%\clearpage

\end{document}